\pdfoutput=1

\documentclass[physrev,superscriptaddress,
reprint
]{revtex4-2}
\usepackage[utf8]{inputenc}
\usepackage{upgreek}
\usepackage{graphicx}
\usepackage{dcolumn}
\usepackage{bm}
\usepackage[english]{babel}
\usepackage{amsmath}
\usepackage{gensymb}
\usepackage[hidelinks]{hyperref}
\usepackage[all]{hypcap}
\usepackage{xcolor} 
\usepackage{soul} 
\usepackage{tabularx}
\usepackage{float}

\begin{document}
\author{Zoltán Kovács-Krausz}
\affiliation{Department of Physics, Institute of Physics, Budapest University of Technology and Economics, Műegyetem rkp. 3., H-1111 Budapest, Hungary}
\affiliation{MTA-BME Superconducting Nanoelectronics Momentum Research Group, Műegyetem rkp. 3., H-1111 Budapest, Hungary}
\author{Endre Tóvári}
\email{tovari.endre@ttk.bme.hu}
\affiliation{Department of Physics, Institute of Physics, Budapest University of Technology and Economics, Műegyetem rkp. 3., H-1111 Budapest, Hungary}
\affiliation{MTA-BME Correlated van der Waals Structures Momentum Research Group, Műegyetem rkp. 3., H-1111 Budapest, Hungary}
\author{Dániel Nagy}
\affiliation{Department of Physics of Complex Systems, ELTE Eötvös Loránd University, 1117 Budapest, Hungary}
\author{Albin Márffy}
\affiliation{Department of Physics, Institute of Physics, Budapest University of Technology and Economics, Műegyetem rkp. 3., H-1111 Budapest, Hungary}
\affiliation{MTA-BME Superconducting Nanoelectronics Momentum Research Group, Műegyetem rkp. 3., H-1111 Budapest, Hungary}
\author{Bogdan Karpiak}
\affiliation{Department of Microtechnology and Nanoscience, Chalmers University of Technology, SE-41296, Göteborg, Sweden}
\author{Zoltán Tajkov}
\affiliation{Centre for Energy Research, Institute of Technical Physics and Materials Science, 1121 Budapest, Hungary}
\author{László Oroszlány}
\affiliation{Department of Physics of Complex Systems, ELTE Eötvös Loránd University, 1117 Budapest, Hungary}
\affiliation{MTA-BME Lendület Topology and Correlation Research Group, Budapest University of Technology and Economics, 1521 Budapest, Hungary}
\author{János Koltai}
\affiliation{ELTE Eötvös Loránd University, Department of Biological Physics, 1117 Budapest, Hungary}
\author{Péter Nemes-Incze}
\affiliation{Centre for Energy Research, Institute of Technical Physics and Materials Science, 1121 Budapest, Hungary}
\author{Saroj Dash}
\affiliation{Department of Microtechnology and Nanoscience, Chalmers University of Technology, SE-41296, Göteborg, Sweden}
\author{Péter Makk}
\affiliation{Department of Physics, Institute of Physics, Budapest University of Technology and Economics, Műegyetem rkp. 3., H-1111 Budapest, Hungary}
\affiliation{MTA-BME Correlated van der Waals Structures Momentum Research Group, Műegyetem rkp. 3., H-1111 Budapest, Hungary}
\author{Szabolcs Csonka}
\affiliation{Department of Physics, Institute of Physics, Budapest University of Technology and Economics, Műegyetem rkp. 3., H-1111 Budapest, Hungary}
\affiliation{MTA-BME Superconducting Nanoelectronics Momentum Research Group, Műegyetem rkp. 3., H-1111 Budapest, Hungary}

\title{Revealing the band structure of ZrTe$_5$ using Multicarrier Transport}

\begin{abstract}

The layered material ZrTe$_5$ appears to exhibit several exotic behaviors which resulted in significant interest recently, although the exact properties are still highly debated. Among these we find a Dirac/Weyl semimetallic behavior, nontrivial spin textures revealed by low temperature transport, and a potential weak or strong topological phase. The anomalous behavior of resistivity has been recently elucidated as originating from band shifting in the electronic structure. Our work examines magnetotransport behavior in ZrTe$_5$ samples in the context of multicarrier transport. The results, in conjunction with ab-initio band structure calculations, indicate that many of the transport features of ZrTe$_5$ across the majority of the temperature range can be adequately explained by the semiclassical multicarrier transport model originating from a complex Fermi surface.

\end{abstract}

\maketitle

\section{Introduction}

The layered quasi-2D material ZrTe$_5$ has attracted considerable attention in the past years due to its unusual transport properties. Experiments suggest the formation of accidental Dirac/Weyl states in the band structure\,\cite{dirac_chen-3Ddirac_msllzs3dmdf_2015,dirac_chen-3DdiracOptical_oss3ddsz_2015,dirac_xiang-diracMagOpt-oq2ddfz_2015,dirac-and-wti_chen-3DDiracIR-sebbi3dmd_2017} leading to semimetallic behavior, as evidenced by chiral magnetic anomaly\,\cite{dirac-chiral_qiang-chiarMagnetic-cmez_2016,dirac-chiral_guolin-diracMagNeg-te3ddspz_2016} and angle-resolved photoemission spectroscopy (ARPES)\,\cite{arpes_moreschini-arpesWTI-ntlesz_2016,bandstruct_manzoni-STIarpes-estipz_2016,arpes_manzoni-ArpesSTI-tdnmbsz_2017,bandstruct-and-fit_zhang-arpeslifshitz-eetilttnz_2017} measurements. These bands are shown to shift significantly with temperature, leading to a characteristic resistivity anomaly in the material, manifesting as a resistance peak at a particular temperature $T_\text{p}$. One complication is that these features depend significantly on the growth method used in ZrTe$_5$ synthesis\,\cite{fit_shahi-bipolar-bcpoetp_2018} - flux and chemical vapor transport (CVT).

\begin{figure*}[!ht]
    \begin{center}
    \includegraphics[width=2.0\columnwidth]{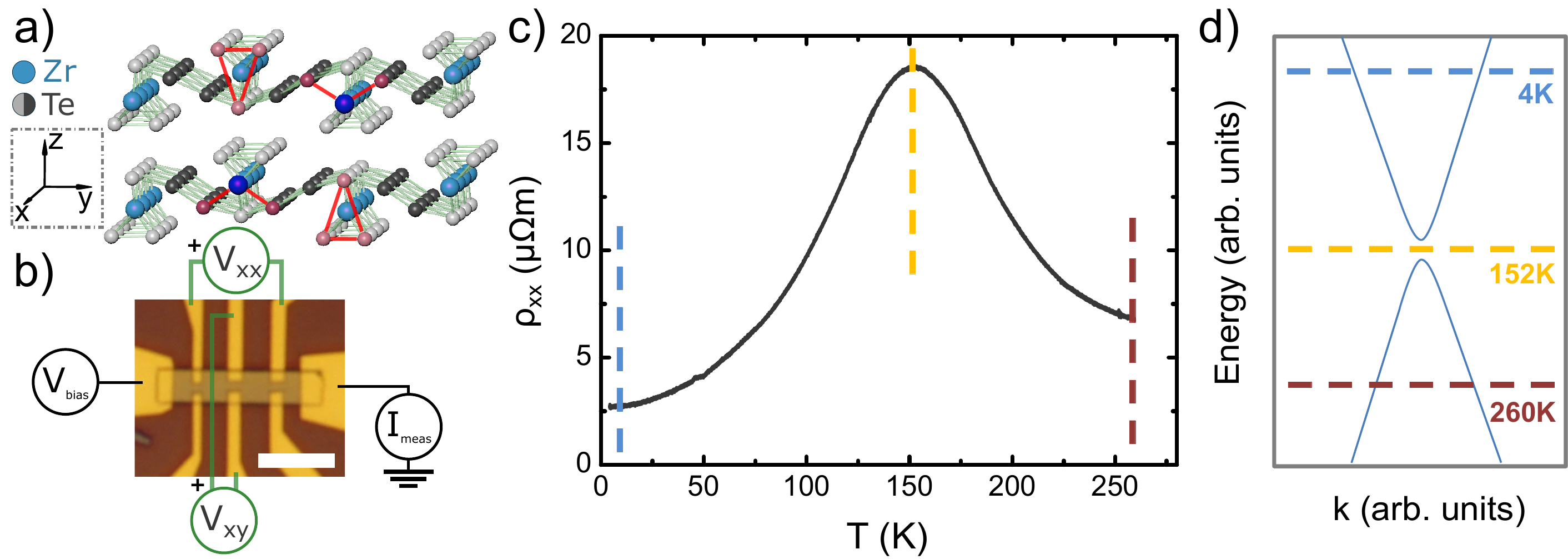}
    \caption{(a) Crystal structure of ZrTe$_5$. The coordinate system is defined such that z is the vdW stacking direction. Components of the unit cell are highlighted with red lines. (b) Exfoliated ZrTe$_5$ crystal on SiO$_2$. The x-axis is along the horizontal direction. The experimental measurement setup is depicted on the Cr/Au electrical contacts deposited on the device. The scale bar is 5 $\upmu$m. (c) Temperature dependence of longitudinal resistivity, exhibiting characteristic peak at $T_\text{p}$=152 K. This behavior is attributed to the Dirac-like bands of ZrTe$_5$ shifting with respect to the chemical potential as the temperature changes. (d) A visual representation of this shift on a model (gapped) Dirac band with the dotted lines representing the points in temperature shown in (c).}
    \label{fig1}
    \end{center}
\end{figure*}

In some works, it is considered to be a topological insulator near the transition between weak and strong topological phases (WTI - STI)\,\cite{bandstruct_fan-WTIvSTI-tbswtizh_2017,dirac-and-sti-and-wti_mutch-strainTunedTransition-esttptz_2019,sdho-torus-flux_wang-magchiralaniso-gmatsz_2022}. This can be reconciled with the previous claim by considering that, near this transition, the energy gap in the Dirac-like bands of the band structure becomes small and transport properties at finite temperature show a Dirac-like behavior. Theoretical band structure calculations\,\cite{bandstruct_weng-QSHparadigm-tmpzh_2014,bandstruct-and-fit_zhang-arpeslifshitz-eetilttnz_2017,bandstruct_fan-WTIvSTI-tbswtizh_2017} based on experimental lattice constants\,\cite{lattice_fjellvag-powDifExp-spzhpd_1986} generally conclude that ZrTe$_5$ is in the STI phase, while being close to the WTI-STI transition, and the outcome is very sensitive to small changes in lattice parameters. Meanwhile, the majority of experimental measurements find evidence of WTI instead\,\cite{bandstruct-and-fit_zhang-arpeslifshitz-eetilttnz_2017,dirac-and-sti-and-wti_mutch-strainTunedTransition-esttptz_2019,bandstruct-dirac_sun-pumpProbeDirac-pdsz_2020,stmEdge_we-stmWTI-etesleg_2016,stmEdge_xbing-stmWTI-eoteds_2016}, and some even describe it to be close enough to the transition point that applying strain, pressure or other external stimuli can lead to the transition to the STI phase\,\cite{bandstruct-and-fit_liu-dynmass-zsdmgdsz_2016,dirac-and-sti-and-wti_mutch-strainTunedTransition-esttptz_2019,wti-sti_tian-bandinversion-debetbi_2019,bandstruct_manzoni-STIarpes-estipz_2016,bandstruct_weng-QSHparadigm-tmpzh_2014,bandstruct-WTIvSTI-tptpsdts_2021}. 

Transport measurements reveal complex magnetoresistance throughout the full temperature range. Some papers treat the magnetotransport features as anomalous Hall effect (AHE) arising from a non-trivial Berry phase in the material\,\cite{bandstruct-and-sdho-pressure_sun-AHEzeeman-lzsiahez_2020,berry-flux_ju-uheibc_2020}. Other works use a multicarrier transport (MCT) model to attempt to explain features such as the simultaneous presence of both types of carriers in the material at some temperatures\,\cite{multiband_tang-3DQHE-3dqhemit_2019,fit_shahi-bipolar-bcpoetp_2018,bandstruct-and-fit_zhang-arpeslifshitz-eetilttnz_2017,bandstruct-and-sdho-pressure_sun-AHEzeeman-lzsiahez_2020,bandstruct-and-fit_liu-dynmass-zsdmgdsz_2016,multiband-and-fit_lu-thicknessDep-tttbtns_2017,multiband-and-sdho_gang_ooeipahmflz_2016,wti-sti_tian-bandinversion-debetbi_2019}. The relatively small energy gap, and the complex band structure with multiple conduction or valence band edges being close to the gap, lends credence to the MCT approach. However, the use of only two distinct carriers at any particular temperature in these works could not fully explain the magnetotransport across the entire temperature range, and particularly near $T_\text{p}$.

In this paper, we demonstrate that the transport behavior of ZrTe$_5$ can indeed be solely described using a semiclassical MCT model across a wide range of temperatures. However, it requires the use of several (up to five) distinct charge carriers, more than what has been used so far in the literature. In addition, we also find that one of these charge carriers is confined to the edge of the sample. Our temperature-dependent magnetotransport measurements are conducted on exfoliated ZrTe$_5$ nanodevices (thickness between 50-150 nm) grown using the CVT method. Though it is possible that factors such as a non-trivial Berry phase also play a role, our results indicate that MCT is able to fully account for the magnetotransport features. Our results are supported by ab-initio band structure calculations. 

\section{Experimental Results}

Devices were fabricated from thin (50-150 nm) exfoliated ZrTe$_5$ crystals by depositing Cr/Au ohmic contacts via electron beam lithography in a geometry depicted in \hyperref[fig1]{Fig.~\ref{fig1}} (b). Before metal deposition, a cleaner contact surface was prepared using an Ar ion beam milling step. Measurements were performed in the temperature ranges of 1.5-305 K with magnetic fields pointing out-of-plane (z-axis on \hyperref[fig1]{Fig.~\ref{fig1}} (a)). Bias voltage was applied along the longitudinal direction (x-axis on \hyperref[fig1]{Fig.~\ref{fig1}} (a)) of the crystals, using the side contacts to simultaneously measure longitudinal ($\rho_\text{xx}$) and transverse ($\rho_\text{xy}$) resistivity in the same crystal. Five similar devices were successfully characterized, here we representatively describe the one shown on \hyperref[fig1]{Fig.~\ref{fig1}} (b) with a thickness of $t=98$ nm as determined by atomic force microscopy. Experimental method details and measurements on further samples are shown in the Supporting Information.

In \hyperref[fig1]{Fig.~\ref{fig1}} (c), the temperature dependence of $\rho_\text{xx}$ is shown at zero magnetic field, featuring the prominent characteristic peak, with $T_\text{p}$ between 130-155 K for all characterized samples. This is typical for ZrTe$_5$ grown with the CVT method\,\cite{dirac-chiral_guolin-diracMagNeg-te3ddspz_2016,multiband_tang-3DQHE-3dqhemit_2019,bandstruct-and-fit_liu-dynmass-zsdmgdsz_2016,fit_shahi-bipolar-bcpoetp_2018}, compared to flux samples\,\cite{dirac-chiral_qiang-chiarMagnetic-cmez_2016,fit_shahi-bipolar-bcpoetp_2018,berry-flux_ju-uheibc_2020,bandstruct-and-sdho-pressure_sun-AHEzeeman-lzsiahez_2020} where it can be as low as 5 K. Recent papers with ARPES measurements\,\cite{bandstruct-and-fit_zhang-arpeslifshitz-eetilttnz_2017} indicate that the resistance peak is due to a shifting of the Dirac-like bands relative to the chemical potential, as the temperature changes (in the following, we denote the chemical potential as $E_\text{F}$, to avoid confusion with carrier mobility). We illustrate this shift in \hyperref[fig1]{Fig.~\ref{fig1}} (d) in conjunction with panel (c), denoting the various positions of $E_\text{F}$ relative to a model Dirac structure with a small energy gap.

\begin{figure}[!ht]
    \begin{center}
    \includegraphics[width=1.0\columnwidth]{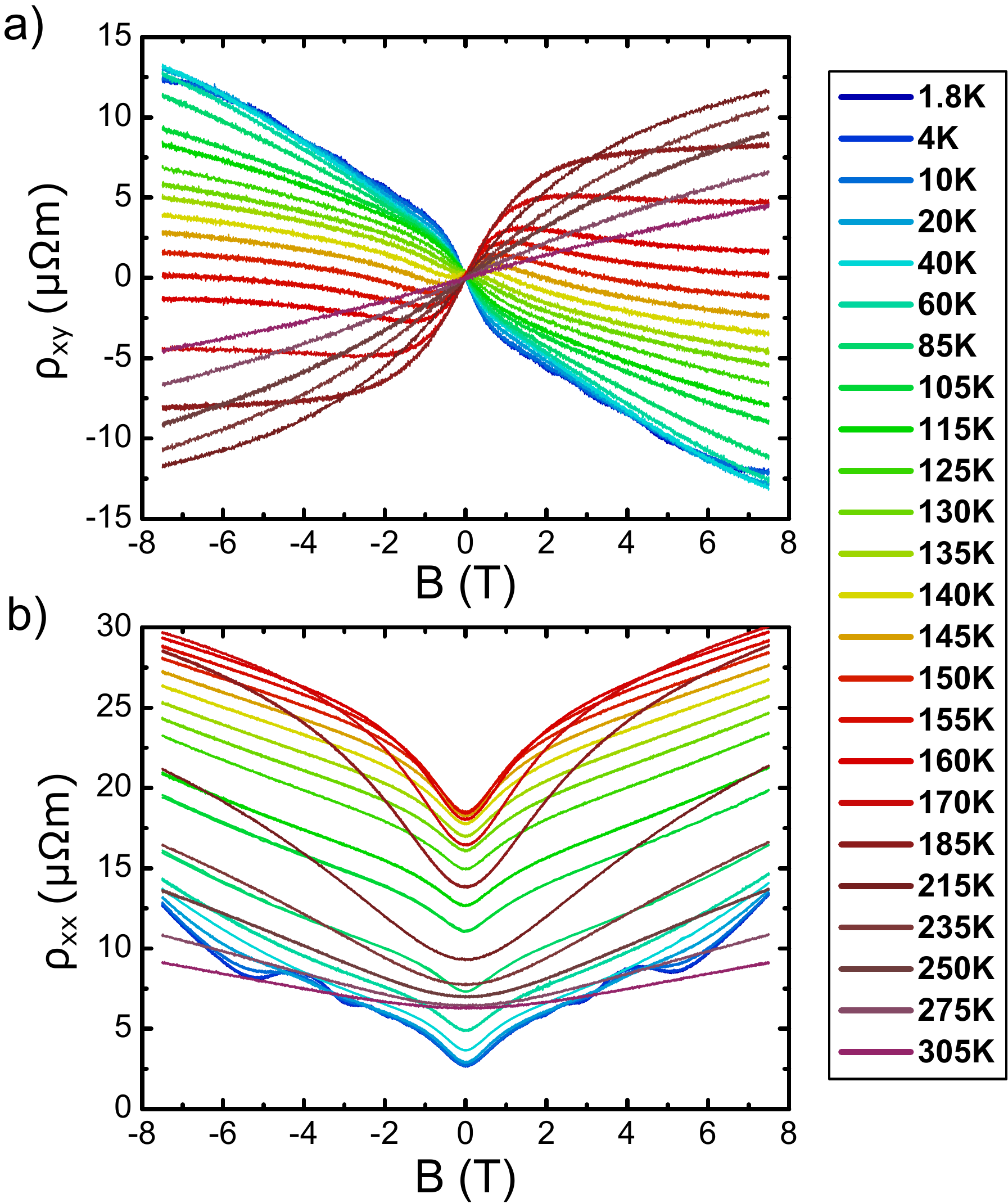}
    \caption{\textbf{Full temperature range magnetotransport in ZrTe$_5$.} (a) Transverse magnetoresistivity and (b) longitudinal magnetoresistivity. The measurements were taken simultaneously at each temperature while sweeping the magnetic field, using the experimental setup in \hyperref[fig1]{Fig.~\ref{fig1}} (b).}
    \label{fig2}
    \end{center}
\end{figure}

The shifting of the bands with respect to $E_\text{F}$ is further supported by Hall measurements: \hyperref[fig2]{Fig.~\ref{fig2}} (a) shows a change in dominant carrier type, visible as the change with temperature of the slope of $\rho_\text{xy}$ near zero magnetic field $B$. However, the full set of measurements across the entire temperature range reveals a much more complex magnetoresistance behavior. Notably, the Hall curves in \hyperref[fig2]{Fig.~\ref{fig2}} (a) show significant nonlinearity at nearly all temperatures. It may be tempting to consider the shapes of some of them as indicative of characteristic AHE-like curves, but as we will show in the following, such curves can also be fully explained using an MCT model.

The $\rho_\text{xx}$ curves exhibit Shubnikov-de Haas oscillations (SdHO) below 20 K (blue curves on \hyperref[fig2]{Fig.~\ref{fig2}} (b)), and large magnetoresistance up to room temperature. The SdHO is often used to obtain the Berry phase using Landau fan diagram analysis. According to most other works\,\cite{dirac_chen-3Ddirac_msllzs3dmdf_2015,dirac-chiral_guolin-diracMagNeg-te3ddspz_2016,sdho-nontrivial_zhang_dadss_2017,sdho-nontrivial_wang_vqods_2018,bandstruct-and-fit_liu-dynmass-zsdmgdsz_2016,sdho-thick_zhuo-monolayer-tdesl2D_2022} the Berry phase from an out-of-plane magnetic field is non-trivial. However, we are able to distinguish multiple frequencies in the SdHO of some devices, where we find that one oscillation corresponds to a non-trivial Berry phase and another to a trivial one. The latter is unusual for ZrTe$_5$, but similar to the findings of Ref.\,\cite{multiband-and-sdho_gang_ooeipahmflz_2016}. Multiple SdHO frequencies have also been noted before\,\cite{SdHO_kamm-multifreq-fsemdt_1985} and indicate multiple Fermi pockets at low temperature, providing further support for MCT. A detailed analysis of the SdHO can be found in the Supporting Information.

\section{Multicarrier Transport}

In the following we will demonstrate that transport features can be well described by the presence of multiple charge carriers contributing to the transport properties. We consider the magnetotransport measurements in the full temperature range in the context of a MCT model with a number of independent carriers (NC) related to separate closed Fermi pockets, each characterized by the carrier type, carrier density and mobility. In such a case (as also seen in e.g. Refs.\,\cite{bandstruct-and-fit_liu-dynmass-zsdmgdsz_2016,multiband_tang-3DQHE-3dqhemit_2019}), the total transverse and longitudinal sheet (2D) conductivity can be written as:

\begin{align}
		\sigma_\text{xy}(B) & = \sum_{\text{i}}^{\text{NC}} \frac{\sigma_\text{i}\mu_\text{i}B}{1+\mu_\text{i}^{2}B^{2}},
	\label{mctcond1}
\end{align}
\begin{align}
		\sigma_\text{xx}(B) & = \sum_{\text{i}}^{\text{NC}} \frac{\lvert \sigma_\text{i} \rvert}{1+\mu_\text{i}^{2}B^{2}},
	\label{mctcond2}
	\end{align}
where $\mu_\text{i}$ and $\sigma_\text{i}$ are the mobility and signed 2D conductivity contribution of the i-th carrier. The conductivity may be written as $\sigma_\text{i}=n_\text{i}q_\text{i}\mu_\text{i}$, where $n_\text{i}$ and $q_\text{i}$ are the 2D carrier density and charge of the corresponding carrier. We have chosen $\mu_\text{i}$ and $\sigma_\text{i}$ as the fit parameters and we have performed simultaneous fits at each temperature of the longitudinal and transverse conductivity, calculated as $\sigma_\text{xx(xy)}=t\rho_\text{xx(xy)}/(\rho_\text{xx}^2 + \rho_\text{xy}^2)$, where $t$ is device thickness, and $\rho_\text{xx(xy)}$ has been (anti-) symmetrized with respect to $B$. For each particular value of NC, a large number of fits was performed using randomized starting parameters, and the results were analyzed in histograms to obtain the most relevant parameter set at each temperature. This analysis was repeated with values of NC from 2 to 6 (see Fig. S4 and S5 of the Supporting Information).

\begin{figure}[!ht]
    \begin{center}
    \includegraphics[width=1.0\columnwidth]{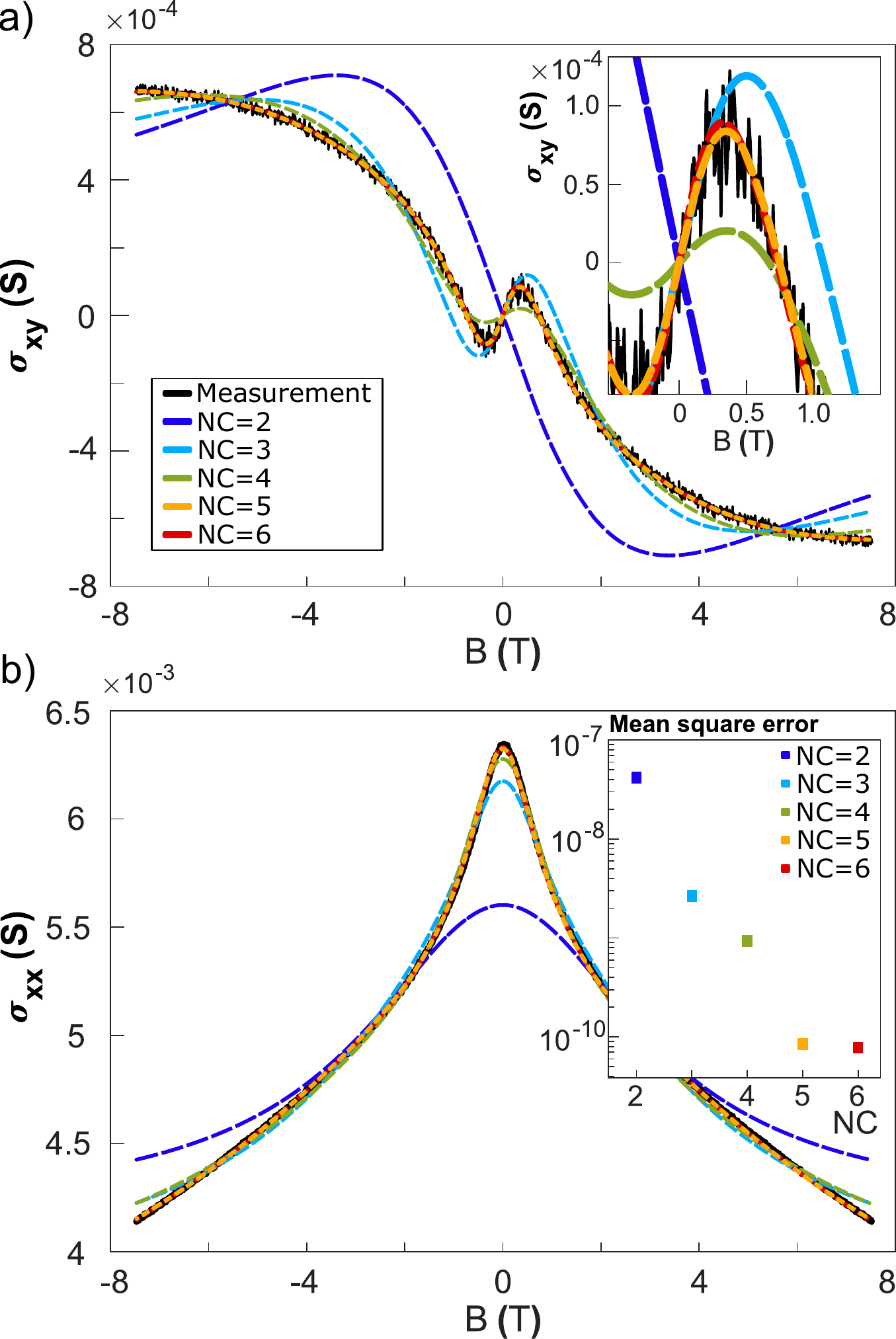}
    \caption{\textbf{Fitting of multiple carriers on magnetotransport data.} Simultaneous fitting of transverse (a) and longitudinal (b) sheet conductivity, with the inset in (a) being magnified detail around the 0 T region of panel (a). The fitting is repeated for different carrier counts (NC) from 2 to 6. The inset in (b) shows the mean square error of the fits. As can be seen on all panels, the fit approaches the experimental curves as NC increases, but reaches an optimal fit at NC=5 and does not improve by further increasing NC (the NC=6 curve is almost completely hidden behind NC=5). The shown example is the fitting of the T=140 K data of the device; the procedure is repeated for all temperatures independently.}
    \label{fig3}
    \end{center}
\end{figure}

\hyperref[fig3]{Fig.~\ref{fig3}} demonstrates the fitting procedure for the T=140 K data curves from \hyperref[fig2]{Fig.~\ref{fig2}}. The number of carriers changes from NC=2 to NC=6, with the fitted transverse and longitudinal conductivity curves (colored curves in \hyperref[fig3]{Fig.~\ref{fig3}} (a) and (b) respectively) providing a progressively closer agreement with the measurement data (in black). In the inset of \hyperref[fig3]{Fig.~\ref{fig3}} (b) the evolution of the mean square error (MSE) associated with various NC values is shown on a logarithmic scale. A marked improvement is shown up to NC=5, while NC=6 shows a negligible improvement, which is also notable on panels (a) and (b) as the two curves visually overlap with each other and the measurement data as well. In comparison, the NC=4 curve still deviates from the measurement data at both low and high magnetic fields. For further details regarding the fitting procedure, consult the Supporting Information.

Unlike previous works, most of which have considered at most two independent carrier contributions at any given temperature, our results suggest that more carriers are necessary to obtain satisfactory agreement with the measurements. Primarily, the increase in NC is justified by the necessity to treat electron and hole contributions separately to account for thermal excitations, especially near $T_\text{p}$. Depending on the temperature, we find that the optimal carrier count varies from NC=3 to NC=5, with the latter number needed to fit the curves close to $T_\text{p}$.

In the following, we analyze the evolution of the fit results with temperature, in order to garner information about the band structure of ZrTe$_5$ from the transport measurements. As we demonstrate in the Supporting Information, we have noticed that, similar to other works using a two-carrier model\,\cite{bandstruct-and-fit_zhang-arpeslifshitz-eetilttnz_2017,fit_shahi-bipolar-bcpoetp_2018,multicar-secondband_chi-lifshitz-ltmetab_2017}, the full temperature range exhibits instabilities for all NC values, particularly in the form of large mobility jumps (sometimes almost two orders of magnitude) within a small temperature range. These discontinuities in mobility are often found in conjunction with the corresponding carrier density dropping to small values and/or a sign change of the carrier. Such drastic mobility discontinuities within small temperature ranges are unlikely to be the result of band shifting caused by temperature. In our datasets, we also find this instability often results in an unreasonably low mobility value for one of the carriers ($\sim 1 \text{cm}^2/\text{Vs}$), while its conductivity is not negligible. The $\sigma_\text{i}=n_\text{i}q_\text{i}\mu_\text{i}$ relation would then imply a huge carrier density, which is not physically reasonable.

\begin{figure}[!ht]
    \begin{center}
    \includegraphics[width=1.0\columnwidth]{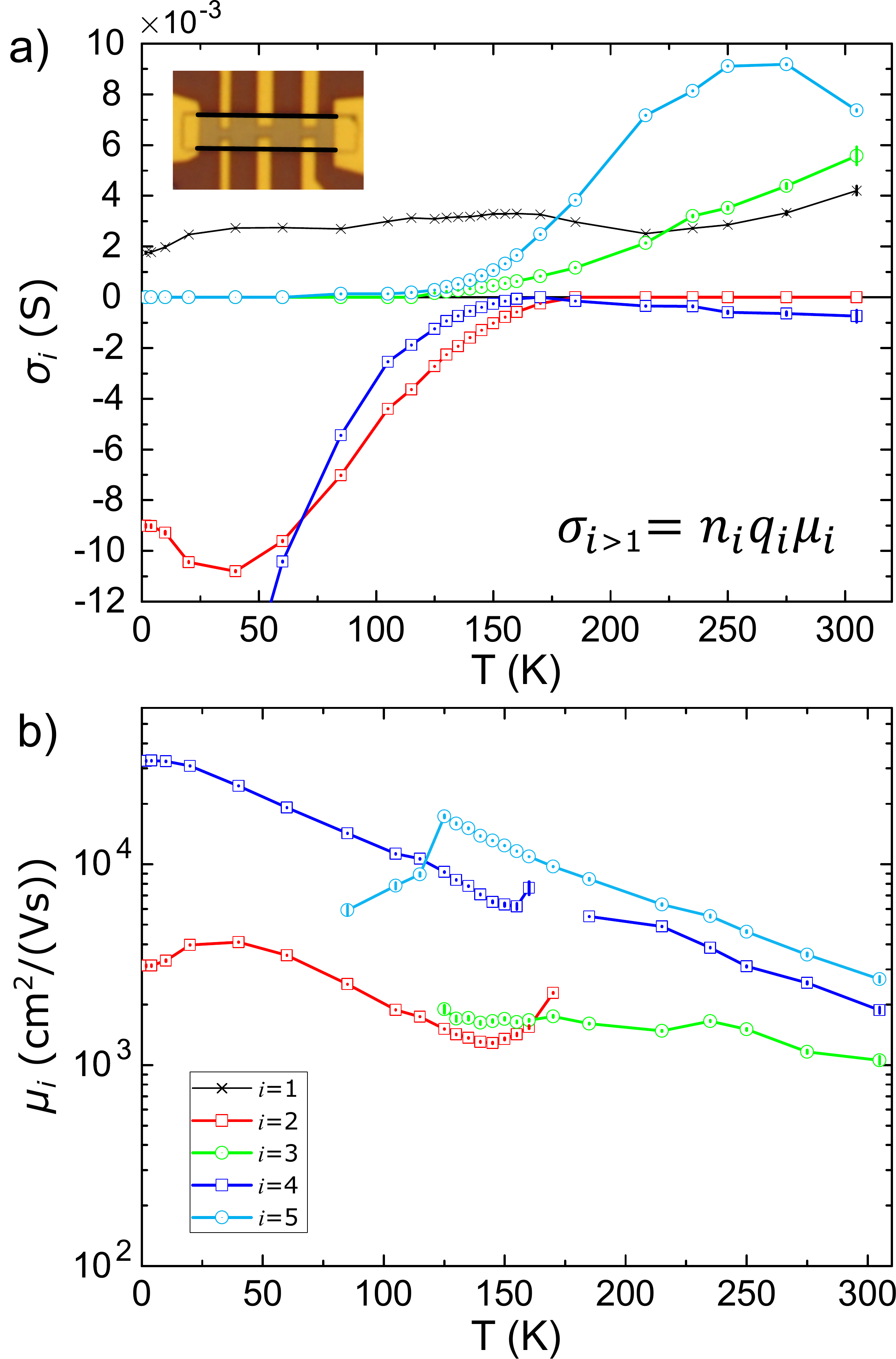}
    \caption{\textbf{Combined results of the multi-carrier fitting.} (a) Signed sheet conductivity contributions of the various obtained subbands as a function of temperature. Their corresponding mobilities are shown in (b). Vertical bars representing the errors of the fits are included. There are a total of 5 subbands identified, although not all play significant roles at every temperature. The subband represented in black (\textit{i}=1) is assumed to be an edge state, as illustrated by black lines in the inset of (a) (hence the lack of $\mu_\text{1}$ in (b)).}
    \label{fig4}
    \end{center}
\end{figure}

Keeping in mind that ZrTe$_5$ is also a candidate topological insulator, we consider the presence of topological edge states in our model and their relevance to the fitting method. In the magnetotransport behavior, the conductivity contribution of an edge state propagating in the xz plane (as in the inset of \hyperref[fig4]{Fig.~\ref{fig4}} (a)) will not depend on a magnetic field along the z-axis. It can be intuitively understood that a diffusive edge-confined carrier would have a finite $\sigma_\text{xx}$ but zero $\sigma_\text{xy}$. In the model (\hyperref[mctcond1]{Eq.~\ref{mctcond1} and Eq.~\ref{mctcond2}}) this can effectively be treated by fixing the mobility of this carrier at zero. This results in no $\sigma_\text{xy}$ contribution, while still allowing a finite, potentially temperature dependent, but field-independent contribution to $\sigma_\text{xx}$ due to finite $\sigma_\text{i}$. The previously mentioned behavior of the fitting, resulting in unreasonably low mobility values for one carrier, can be interpreted in this context as the effect of an edge state's contribution.

Therefore we fixed the $\sigma_\text{xx}$ contribution of one carrier to a field-independent $|\sigma_\text{i}|$, and its $\sigma_\text{xy}$ contribution to zero, and repeated the fitting procedure. The previously seen issue of mobility discontinuities with temperature was resolved, as long as the NC value was not too small (see Fig. S5 of Supporting Information). We note that the fits shown in \hyperref[fig3]{Fig.~\ref{fig3}} had already been performed this way. The results of the fitting procedure, in the form of signed conductivity and corresponding mobility values of each particular carrier in the model, are shown in \hyperref[fig4]{Fig.~\ref{fig4}} (a) and (b) respectively, as a function of temperature. The carriers of the model are color-coded, and circles in the dataset represent electron bands while squares represent hole bands. The carrier type is determined by the sign of $\sigma_\text{i}$ (owing to $q_\text{i}$). The black datapoints of \hyperref[fig4]{Fig.~\ref{fig4}} (a) represent the conductivity contribution of the edge state, which is consistent over the full temperature range, and has no corresponding mobility information. In the Supporting Information (Fig. S2) we show that the fitting results in all characterized devices consistently demonstrate the same MCT features.

\section{Modeling and Discussion}

Of the five charge carriers in \hyperref[fig4]{Fig.~\ref{fig4}}, the four bulk carriers (\textit{i}=2-5) can be grouped into a pair with comparatively similar high mobility (blue and teal) and a pair with lower mobility (red and green). There is one electron and one hole type carrier in each pair. We attribute the higher mobility carriers to the Dirac-like bands of ZrTe$_5$, where the downward shift of the bands relative to $E_\text{F}$ (as temperature decreases) leads to a change from hole to electron-like carriers, as seen in \hyperref[fig1]{Fig.~\ref{fig1}} (d). This result is similar to previously considered illustrative band models in other works, which tend to feature the Dirac band as well as a conventional side-band\,\cite{bandstruct-and-fit_liu-dynmass-zsdmgdsz_2016,multiband-and-fit_lu-thicknessDep-tttbtns_2017,multicar-secondband_chi-lifshitz-ltmetab_2017,multiband-nano_jj-etnzs_2017}. This side-band is either only represented as a conduction band alone, similar to the electron pockets revealed via ARPES in Ref.\,\cite{bandstruct-and-fit_zhang-arpeslifshitz-eetilttnz_2017}, or as a pair of conduction and valence bands. In our model, all four bands contribute to transport throughout large parts of the temperature range, and particularly at the temperatures near $T_\text{p}$. As the temperature decreases, the holes in the valence bands (green and teal bands of \hyperref[fig4]{Fig.~\ref{fig4}} (a)) effectively freeze out at low temperature, while at high temperature the electron bands (red and blue) are depleted - although not completely emptied, due to the presence of thermal excitations. This suggests that, as in Ref.\,\cite{bandstruct-and-fit_zhang-arpeslifshitz-eetilttnz_2017}, all bands shift downward in energy with respect to $E_\text{F}$ as the temperature decreases. We note that while near $T_\text{p}$, five carriers were required for a good fit (as in \hyperref[fig3]{Fig.~\ref{fig3}}), further from the peak four or even three carriers were adequate. These changes in NC do not lead to discrepancies in the trends of \hyperref[fig4]{Fig.~\ref{fig4}} apart from small discontinuities in mobility near $\sigma_\text{i} \propto n_\text{i} \approx 0$, and are consistent with the above described expectation in depletion of carriers. Moreover, the mobilities generally decrease with increasing temperature, as expected from an increase in scattering rates (e.g. electron-phonon scattering). The lowest temperature datapoints should not be considered accurate, since here Landau level physics and phase coherence effects likely introduce artifacts into the fitting results.

\begin{figure*}[!ht]
    \begin{center}
    \includegraphics[width=2.0\columnwidth]{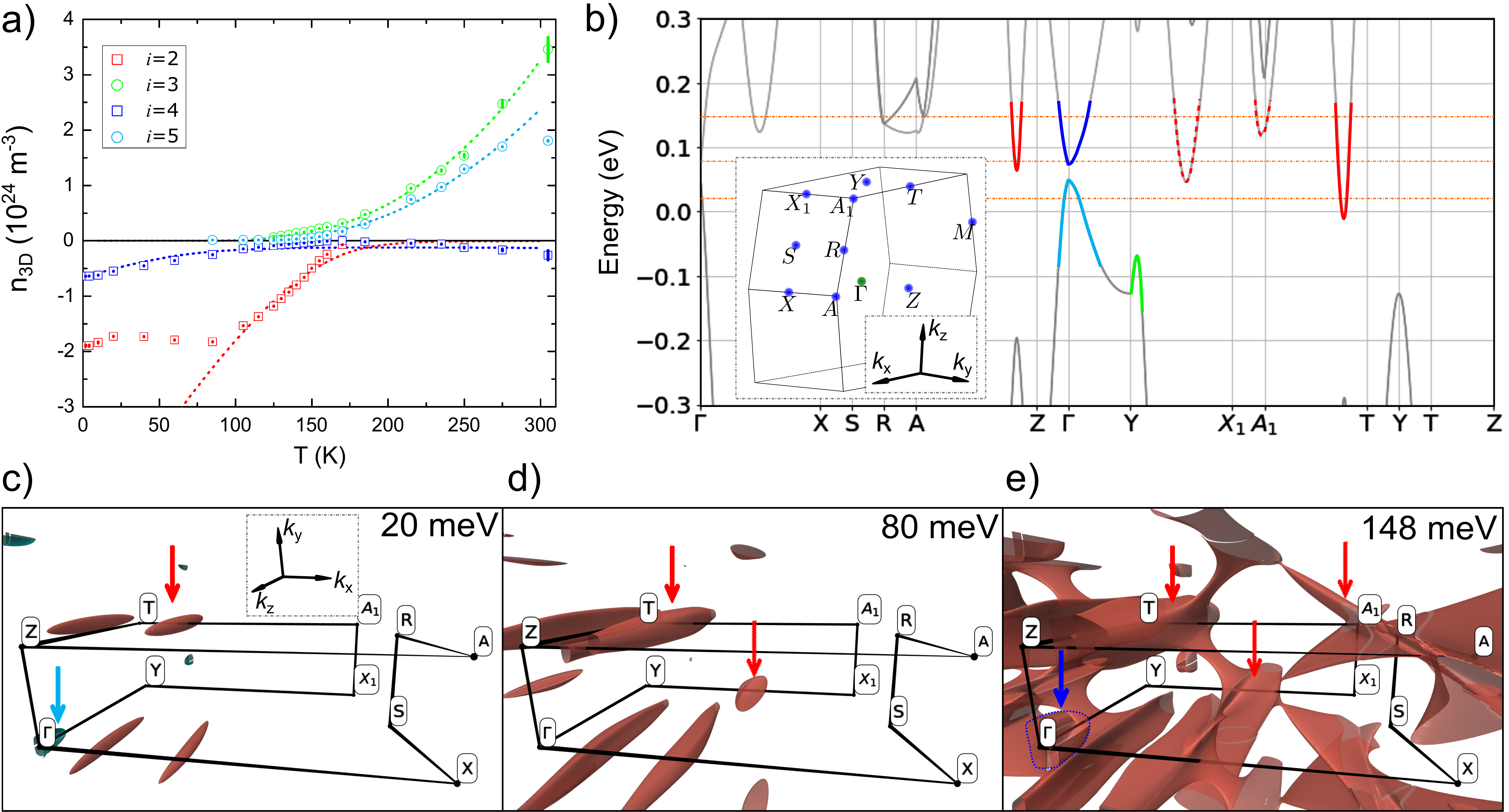}
    \caption{(a) Signed carrier densities of the various carriers in the MCT model (discrete datapoints) and fit results (dotted lines) for each carrier (parabolic dispersion for $i=2,3$ and Dirac for $i=4,5$). The color notations are identical to Fig. 4, and error bars are included. (b) DFT results of the first conduction and valence bands of ZrTe$_5$. Color notations on the band structure show the best guess correspondence to the carriers from (a). (c)-(e) Energy isosurfaces at $E_\text{F} =$ 20, 80 and 148 meV, corresponding to orange dashed lines of panel (b). The path in black follows the x-axis in panel (b) along high symmetry points of the BZ. The isosurface color is blue for the valence band and red for the conduction band. The colored arrows denote the features of the band structure in (b) depicted in the same colors.}
    \label{fig5}
    \end{center}
\end{figure*}

The temperature dependence of individual signed carrier densities, $n_\text{3D,i}(T)$, is shown as discrete datapoints in \hyperref[fig5]{Fig.~\ref{fig5}} (a), obtained by dividing $\sigma_\text{i}$ with $\mu_\text{i}e$ in the data from \hyperref[fig4]{Fig.~\ref{fig4}}, and further by the sample thickness of 98 nm. The change of carrier densities with temperature is assumed to originate from the shifting bands with respect to $E_\text{F}$ as well as due to the temperature-dependent Fermi function affecting the filling of the bands.

In the following, we obtain band structure information from the temperature dependence of the carrier densities, and compare them with density functional theory (DFT) calculations detailed further below. The carrier density at a particular temperature is the integral of the density of states (DOS), $g$, of the respective band and the Fermi function, $f$, over the entire energy range. If we assume each carrier to correspond to a simple parabolic or Dirac band in the band structure, the DOS depends on the band parameters: band edge energy $E_\text{i}$ and effective mass $m^*_\text{i}$ (or Fermi velocity $v_\text{F,i}$ in case of Dirac bands). Note that, in the picture of the bands shifting relative to $E_\text{F}$, the band edge energies $E_\text{i}$ refer to their position at room temperature $T_0=300$ K, relative to the reference energy level ($\varepsilon=E_\text{F}=0$), and we assume that the bands shift rigidly with the same rate as the temperature changes. The carrier density can be written as:
\begin{align}
		n_\text{3D,i}(T)=\int_{\varepsilon}g(\varepsilon,E_\text{i},m^*_\text{i})f(\varepsilon,T,r_\text{i})d\varepsilon.
	\label{ntintegral1}
\end{align}
Here the parameter $r_\text{i}$ is the shifting rate of the band structure relative to $E_\text{F}$. A starting value for this rate (of approximately 0.25 meV/K) can be extracted from ARPES results performed on a similar CVT-grown sample with $T_\text{p}=135$ K\,\cite{bandstruct-and-fit_zhang-arpeslifshitz-eetilttnz_2017}. We have incorporated this parameter into a modified form for the Fermi function:
\begin{equation}
\begin{aligned}
		f(\varepsilon,T,r_\text{i}) = \frac{1}{1+e^{\frac{\varepsilon-\left [ E_\text{F} + r_\text{i}\left ( T_0-T \right ) \right ]}{kT}}} d\varepsilon,
	\label{ntintegralFermi}
\end{aligned}
\end{equation}
The carrier density expression for e.g. an electron pocket with 3D parabolic dispersion, is the following:
\begin{equation}
\begin{aligned}
		n_\text{3D,i}(T) & =-\frac{1}{2\pi^2}\sqrt{\left ( \frac{2m^*_\text{i}}{\hbar^2} \right )^3} \int_{\varepsilon} \sqrt{\varepsilon-E_\text{i}} f(\varepsilon,T,r_\text{i}) d\varepsilon,
	\label{ntintegral2}
\end{aligned}
\end{equation}
while for hole type carriers the sign is positive, the DOS uses $\sqrt{E_\text{i}-\varepsilon}$ and the Fermi function is replaced with $1-f(\varepsilon,T,r_\text{i})$. Conversely, for the 3D Dirac-like carriers the DOS is $\frac{|E_\text{i}-\varepsilon|^2}{2\pi^2\hbar^3v_\text{F,i}^3}$.

By fitting \hyperref[ntintegral1]{Eq.~\ref{ntintegral1}} on the $n_\text{3D,i}(T)$ data from \hyperref[fig5]{Fig.~\ref{fig5}} (a), we are able to obtain the effective band parameters associated with each carrier involved in the transport. The fit results for the carrier density are shown as dotted lines on the same plot, while the obtained band parameters are described in \hyperref[bandcoeftable]{Table ~\ref{bandcoeftable}}. Note that, for the red ($i=2$) carrier, only points above 100 K were used for the fit (as detailed further below). The blue and teal ($i=4,5$; high mobility) bands are assumed to be the 3D Dirac-like bands of ZrTe$_5$ near the $\Upgamma$ point of the Brillouin zone (BZ), see \hyperref[fig5]{Fig.~\ref{fig5}} (b). The difference between the electron and hole band edge energies ($E_4-E_5$) should be indicative of the direct bandgap of the band structure near $\Upgamma$, hereafter $\Delta_\Upgamma$. These band edge energies are $E_5 \approx 4.6$ meV for the holes, and $E_4 \approx 18.4$ meV for the electrons, resulting in a gap $\Delta_\Upgamma=13.8 \pm 15.0$ meV. For comparison, using a 3D parabolic DOS instead, due to the sharper cutoff of the DOS the resulting gap is slightly larger: $\Delta_\Upgamma=28 \pm 11$ meV. We also mention that, if the material were quasi-2D Dirac, with linear dispersion in-plane and quadratic in the z-axis (see Refs.\,\cite{quasi2D_yang-quantOsc-qonsq2dds-2019,multiband_tang-3DQHE-3dqhemit_2019}), or if it had a gapped 3D Dirac dispersion, we would expect $\Delta_\Upgamma$ to fall in between these values.

\begin{table}
\begin{tabularx}{\columnwidth}{|X|c|c|c|X|}
\hline
Carrier & DOS & $E_\text{i}$ & $m^*_\text{i}$ or $v_\text{F,i}$ & $r_\text{i}$ \\
$i$, type & & [meV] & [$m_\text{e}$ or $10^4\frac{m}{s}$] & [meV/K]\\
\hline
\textcolor[rgb]{1,0,0}{2, $e^-$} & 3D parabolic & 92 $\pm$ 9 & 0.13 $\pm$ 0.03 & 0.66 $\pm$ 0.28 \\
\hline
\textcolor[rgb]{0,1,0}{3, $h^+$} & 3D parabolic & -62.6 $\pm$ 1.3 & 2.177 $\pm$ 0.073 & 0.25* \\
\hline
\textcolor[rgb]{0,0,1}{4, $e^-$} & 3D Dirac & 18.4 $\pm$ 11.7 & 21.0 $\pm$ 4.6 & 0.25* \\
\, & 3D parabolic & 42 $\pm$ 10 & 0.13 $\pm$ 0.04 & 0.25* \\
\hline
\textcolor[rgb]{0,0.68,0.94}{5, $h^+$} & 3D Dirac & 4.6 $\pm$ 3.3 & 16.0 $\pm$ 0.7 & 0.25* \\
\, & 3D parabolic & 14 $\pm$ 1 & 0.33 $\pm$ 0.01 & 0.25* \\
\hline
\end{tabularx}
\caption{Band parameters obtained by fitting the $n_\text{3D,i}(T)$ data of \hyperref[fig5]{Fig.~\ref{fig5}} (a). For the Dirac-like bands ($i=$4,5), in addition to the 3D Dirac DOS, the parabolic case was also used for comparison. For the band shift rates marked with *, using $r_\text{i}$ as a free parameter did not improve the fit result, and it was instead fixed to the starting value.}
\label{bandcoeftable}
\end{table}

To compare these fit results to the theoretical electronic structure of ZrTe$_5$, we performed ab-initio band structure calculations using the SIESTA implementation of DFT. Compared to the standard lattice parameters based on Ref.\,\cite{lattice_fjellvag-powDifExp-spzhpd_1986} and experimentally verified by X-ray diffraction ($a=1.994$ \AA, $b=7.265$ \AA, $c=13.724$ \AA), the relaxed lattice parameters from DFT deviate by about 1\% ($a=2.002$ \AA, $b=7.204$ \AA, $c=13.876$ \AA).\,\cite{topo_tajkov-phase-strain_2022}. Using these parameters, the DFT results (shown in Fig. S7 (a) of the Supporting Information) predict $\Delta_\Upgamma$ to be much larger ($\sim$80 meV) than our fit results. Moreover, we consider the hole sideband along Y-X$_1$ in the band structure, which we attribute to carrier $i=3$. The DFT result puts its maximum at approximately -20 meV instead of the -63 meV we obtain from the fitting above. In addition, the material is predicted to be in the STI phase.

While our transport measurements cannot distinguish the WTI and STI phases, the smaller direct gap obtained by our approach implies the material is closer to the topological phase transition than predicted by DFT. Experimental works using ARPES usually also find a smaller energy gap than theoretically predicted, and often find that the material appears to be in the WTI phase,\,\cite{bandstruct-and-fit_zhang-arpeslifshitz-eetilttnz_2017,bandstruct_manzoni-STIarpes-estipz_2016,bandstruct-and-sdho-pressure_sun-AHEzeeman-lzsiahez_2020,bandstruct-dirac_sun-pumpProbeDirac-pdsz_2020,dirac-and-wti_chen-3DDiracIR-sebbi3dmd_2017} both of which require slightly different lattice constants for the DFT method to reproduce\,\cite{topo_tajkov-phase-strain_2022}. One possible explanation for this disparity is that the CVT method is known for producing Te vacancies in the material\,\cite{fit_shahi-bipolar-bcpoetp_2018}, which leads to n-doping and the shifting of $T_\text{p}$ above 100 K. Theoretical works, especially band structure calculations, rarely take vacancies into account. 

While the experimental energy gap can be reproduced by the DFT results solely by changing the interlayer spacing, as suggested in Ref.\,\cite{bandstruct-and-fit_zhang-arpeslifshitz-eetilttnz_2017}, this requires an up to 8-10\% expansion of the lattice parameter $b$. Instead, we consider an isotropic increase of approximately 2\% in all lattice parameters (similar to Refs.\,\cite{bandstruct_weng-QSHparadigm-tmpzh_2014,bandstruct_fan-WTIvSTI-tbswtizh_2017}). This choice of lattice parameters is a more modest deviation from the diffraction measurements, and comparable to the 1\% deviation of the relaxed parameters of the DFT method. Note that an isotropic modification can be seen as the DFT method underestimating the unit cell size. We find that the DFT results using these lattice parameters better reproduce the -63 meV energy of the $i=3$ (green) hole sideband, in addition to the size of the direct gap near $\Upgamma$, and the material is also predicted to be in WTI phase. The resulting calculated band structure is shown in \hyperref[fig5]{Fig.~\ref{fig5}} (b), on which we have highlighted the best candidate bands that correspond to those obtained from the experimental fits, using the same colors. The band edge energies on the DFT result are higher than the fitted parameters, but this might be explained by the inherent n-doping of CVT-grown ZrTe$_5$, which is not represented in the theoretical calculation.

The remaining carrier in the MCT model not discussed so far is that of the low-mobility electron sideband ($i=2$, red data on \hyperref[fig5]{Fig.~\ref{fig5}}). Its carrier density is difficult to fit with a simple 3D parabolic band, and the fit is reasonable only in the 100 - 200 K range. Using the starting value of 0.25 meV/K for the band shift rate, above 200 K we expect significant remnant carrier density owing to thermally excited electrons, while below 100 K the model predicts a continued increase in carrier density as the band shifts further downwards in energy compared to $E_\text{F}$. Instead, the data shows vanishing density at high temperature and a saturation and even depletion at low temperature. When the fitting is performed on the data in the 100 - 200 K temperature range, as seen in \hyperref[fig5]{Fig.~\ref{fig5}} (a), by using the band shift rate as a free fitting parameter instead of a fixed one like before, the obtained shift rate is significantly higher ($0.66 \pm 0.28$ meV/K) than the 0.25 meV/K starting value, suggesting that this band would have to shift at a much larger rate than the others. It is noteworthy that, of the highlighted red electron band candidates in \hyperref[fig5]{Fig.~\ref{fig5}} (b), the one along A$_1$-T is experimentally detected as an electron pocket in ARPES studies\,\cite{bandstruct-and-fit_zhang-arpeslifshitz-eetilttnz_2017,sideband-epocket_shen-segesb_2017,stmEdge_we-stmWTI-etesleg_2016}. In that study, this electron sideband does indeed appear to shift at an increased rate (approx. 0.45 meV/K) with temperature. However, our obtained band edge energy of around 90 meV for the $i=2$ data (see \hyperref[bandcoeftable]{Table ~\ref{bandcoeftable}}) does not closely match any of the candidate electron pockets on the band structure.

Examining the energy isosurfaces through the entire BZ (\hyperref[fig5]{Fig.~\ref{fig5}} (c) to (e)) reveals an explanation for the inability to fit the $i=2$ electron sideband, as well as a limitation to our MCT approach. For $E_\text{F}=20$ meV energy, a single electron-type ellipsoid surface (multiplied by lattice symmetries) is found, along A$_1$-T (highlighted by a red arrow \hyperref[fig5]{Fig.~\ref{fig5}} (c)), along with the separate hole pocket on $\Upgamma$-Y (highlighted by a teal arrow), attributed to the $i=5$ carrier. However, as can be seen in \hyperref[fig5]{Fig.~\ref{fig5}} (b) along the Y-X$_1$ line, another electron pocket appears at 48 meV. Intuitively, this pocket could also appear in the MCT model as an additional carrier; but increasing the number of fitted carriers (NC=6) did not result in the identification of such a pocket. If the carriers associated with the two pockets have similar mobilities, then the MCT approach cannot distinguish between them, and they will both appear as a single mixed carrier contribution in the NC=5 case. Fitting \hyperref[ntintegral2]{Eq.~\ref{ntintegral2}} would not work here as two electron pockets are not described by the DOS of one parabolic band.

At $E_\text{F}=80$ meV (\hyperref[fig5]{Fig.~\ref{fig5}} (d)) we can see a significant Y-X$_1$ pocket. Additionally, the Fermi surface corresponding to the previous A$_1$-T pocket has grown along the $k_\text{z}$ direction across the entire BZ, and can no longer be described by a closed ellipsoid. At 148 meV (\hyperref[fig5]{Fig.~\ref{fig5}} (e)) the landscape is even more distorted, with the X$_1$-A$_1$ pocket also appearing as oblique lobes, joining with the surfaces of the previous two pockets and forming one large connected surface which is now open along both $k_\text{y}$ and $k_\text{z}$. Note that the $\Upgamma$-X band (in \hyperref[fig5]{Fig.~\ref{fig5}} (b)) belongs to the same pocket as Y-X$_1$, while the band along R-A belongs to the same pocket as X$_1$-A$_1$; therefore these bands have not been highlighted on the band structure. The separate electron pocket corresponding to the Dirac-like band at $\Upgamma$ has been highlighted in blue (the $i=4$ carrier). Taking into consideration the band shift of 75 meV over the full temperature range (using 0.25 meV/K) and the inherent n-doping of CVT-grown ZrTe$_5$, this open Fermi surface above 100 meV can contribute to transport at low temperatures, making \hyperref[mctcond1]{Eq.~\ref{mctcond1}} and \hyperref[mctcond2]{Eq.~\ref{mctcond2}} of the MCT model, and thus the $\sigma_2, \mu_2$ fitting results, inadequate.

Altogether, for most of the studied temperature range, the overall results of our MCT modeling approach are consistent with the theoretical band structure regarding the involvement of multiple band edges in the transport characteristics. Near $T_\text{p}$ up to five independent carriers were required to obtain adequate fits to the transport measurements: two Dirac-like high mobility carriers, two lower mobility carriers, and one carrier confined to the edges. Further away from $T_\text{p}$, the required number of carriers is reduced to four or sometimes three. This behavior is consistent with the depopulation of transport bands as their edges move further away from the chemical potential with temperature, and can be traced in the model as a vanishing of the conductivity contribution and carrier density of the carrier in question. Due to extremely low mobility behavior, one of these carriers is assumed to be located on the edges of the ZrTe$_5$ crystal. This might originate, for example, in an oxidized crystal edge contributing parallel conduction, or a topological edge state.

In conclusion, we have used a MCT model to investigate transport features of ZrTe$_5$. We find that our approach can explain most of the peculiar transport features of the material. Near $T_\text{p}$ an unprecedented number of carriers (NC=5) is required to adequately explain the measurements, in all our measured devices. We confirm the presence of a separate conducting channel localized on the edges of ZrTe$_5$. We are able to identify some carriers as features of the band structure obtained from DFT calculations. This detailed analysis of transport measurements, if combined with other methods such as ARPES, could reveal further details of the ZrTe$_5$ band structure and support the development of more accurate theoretical models. As the band structure is highly sensitive to lattice parameters, one avenue for further research could be modifying them with e.g. hydrostatic pressure and analyzing the changes using MCT. We have recently reported a practical in-situ measurement method for nanodevices under pressure\,\cite{pressure_fulop-vdW-pressure_2021}.

\section*{Author Contributions}
Z.K.K. and B.K. fabricated the devices, with help from A.M.. Transport measurements were performed by Z.K.K., B.K. and A.M.. Transport data analysis and fitting was performed by Z.K.K., E.T., P.M. and S.C.. DFT calculations were performed by D.N., Z.T., L.O. and J.K.. All authors contributed to the manuscript and discussions. S.C., S.P.D., P.M., N.I.P. and E.T. planned and guided the project.

\section*{Acknowledgments}
The authors are thankful to the Institute of Technical Physics and Materials Science of the Centre for Energy Research for providing their facilities for sample fabrication, and M. G. Beckerné, F. Fülöp, M. Hajdu for their technical support. This work has received funding from the FlagERA MultiSpin network, OTKA grants K-138433, K-134437, K-131938, FK-123894, FK-124723 and PD-134758, the FET Open AndQC network with GA number 828948, and the Bolyai Scholarship of the Hungarian Academy of Sciences (Grant BO/00242/20/11). This research was supported by the Ministry of Innovation and Technology and the National Research, Development and Innovation Office within the Quantum Information National Laboratory of Hungary. We acknowledge COST Action CA 21144 SUPERQUMAP, and support from the Graphene Flagship and from the European Union’s Horizon 2020 research and innovation program under Grant Agreement 862660/QUANTUM E LEAPS. L.O. also acknowledges support of the National Research, Development and Innovation (NRDI) Office of Hungary and the Hungarian Academy of Sciences through the Bolyai and Bolyai+ scholarships. P.N.I., Z.T. and S.P.D. acknowledge support from the Graphene Flagship (Core 3, No. 881603). S.P.D. further acknowledges funding 2D TECH VINNOVA center (No. 2019-00068), Swedish Research Council VR project grants (No. 2021–04821) and FLAG-ERA project 2DSOTECH (VR No. 2021-05925). Low T infrastructure was provided by VEKOP-2.3.3-15-2017-00015.


\begin{thebibliography}{0}%
\makeatletter
\providecommand \@ifxundefined [1]{%
 \@ifx{#1\undefined}
}%
\providecommand \@ifnum [1]{%
 \ifnum #1\expandafter \@firstoftwo
 \else \expandafter \@secondoftwo
 \fi
}%
\providecommand \@ifx [1]{%
 \ifx #1\expandafter \@firstoftwo
 \else \expandafter \@secondoftwo
 \fi
}%
\providecommand \natexlab [1]{#1}%
\providecommand \enquote  [1]{``#1''}%
\providecommand \bibnamefont  [1]{#1}%
\providecommand \bibfnamefont [1]{#1}%
\providecommand \citenamefont [1]{#1}%
\providecommand \href@noop [0]{\@secondoftwo}%
\providecommand \href [0]{\begingroup \@sanitize@url \@href}%
\providecommand \@href[1]{\@@startlink{#1}\@@href}%
\providecommand \@@href[1]{\endgroup#1\@@endlink}%
\providecommand \@sanitize@url [0]{\catcode `\\12\catcode `\$12\catcode
  `\&12\catcode `\#12\catcode `\^12\catcode `\_12\catcode `\%12\relax}%
\providecommand \@@startlink[1]{}%
\providecommand \@@endlink[0]{}%
\providecommand \url  [0]{\begingroup\@sanitize@url \@url }%
\providecommand \@url [1]{\endgroup\@href {#1}{\urlprefix }}%
\providecommand \urlprefix  [0]{URL }%
\providecommand \Eprint [0]{\href }%
\providecommand \doibase [0]{https://doi.org/}%
\providecommand \selectlanguage [0]{\@gobble}%
\providecommand \bibinfo  [0]{\@secondoftwo}%
\providecommand \bibfield  [0]{\@secondoftwo}%
\providecommand \translation [1]{[#1]}%
\providecommand \BibitemOpen [0]{}%
\providecommand \bibitemStop [0]{}%
\providecommand \bibitemNoStop [0]{.\EOS\space}%
\providecommand \EOS [0]{\spacefactor3000\relax}%
\providecommand \BibitemShut  [1]{\csname bibitem#1\endcsname}%
\let\auto@bib@innerbib\@empty
\end{thebibliography}%


\begin{thebibliography}{37}%
\makeatletter
\providecommand \@ifxundefined [1]{%
 \@ifx{#1\undefined}
}%
\providecommand \@ifnum [1]{%
 \ifnum #1\expandafter \@firstoftwo
 \else \expandafter \@secondoftwo
 \fi
}%
\providecommand \@ifx [1]{%
 \ifx #1\expandafter \@firstoftwo
 \else \expandafter \@secondoftwo
 \fi
}%
\providecommand \natexlab [1]{#1}%
\providecommand \enquote  [1]{``#1''}%
\providecommand \bibnamefont  [1]{#1}%
\providecommand \bibfnamefont [1]{#1}%
\providecommand \citenamefont [1]{#1}%
\providecommand \href@noop [0]{\@secondoftwo}%
\providecommand \href [0]{\begingroup \@sanitize@url \@href}%
\providecommand \@href[1]{\@@startlink{#1}\@@href}%
\providecommand \@@href[1]{\endgroup#1\@@endlink}%
\providecommand \@sanitize@url [0]{\catcode `\\12\catcode `\$12\catcode
  `\&12\catcode `\#12\catcode `\^12\catcode `\_12\catcode `\%12\relax}%
\providecommand \@@startlink[1]{}%
\providecommand \@@endlink[0]{}%
\providecommand \url  [0]{\begingroup\@sanitize@url \@url }%
\providecommand \@url [1]{\endgroup\@href {#1}{\urlprefix }}%
\providecommand \urlprefix  [0]{URL }%
\providecommand \Eprint [0]{\href }%
\providecommand \doibase [0]{https://doi.org/}%
\providecommand \selectlanguage [0]{\@gobble}%
\providecommand \bibinfo  [0]{\@secondoftwo}%
\providecommand \bibfield  [0]{\@secondoftwo}%
\providecommand \translation [1]{[#1]}%
\providecommand \BibitemOpen [0]{}%
\providecommand \bibitemStop [0]{}%
\providecommand \bibitemNoStop [0]{.\EOS\space}%
\providecommand \EOS [0]{\spacefactor3000\relax}%
\providecommand \BibitemShut  [1]{\csname bibitem#1\endcsname}%
\let\auto@bib@innerbib\@empty
\bibitem [{\citenamefont {Chen}\ \emph
  {et~al.}(2015{\natexlab{a}})\citenamefont {Chen}, \citenamefont {Chen},
  \citenamefont {Song}, \citenamefont {Schneeloch}, \citenamefont {Gu},
  \citenamefont {Wang},\ and\ \citenamefont
  {Wang}}]{dirac_chen-3Ddirac_msllzs3dmdf_2015}%
  \BibitemOpen
  \bibfield  {author} {\bibinfo {author} {\bibfnamefont {R.~Y.}\ \bibnamefont
  {Chen}}, \bibinfo {author} {\bibfnamefont {Z.~G.}\ \bibnamefont {Chen}},
  \bibinfo {author} {\bibfnamefont {X.-Y.}\ \bibnamefont {Song}}, \bibinfo
  {author} {\bibfnamefont {J.~A.}\ \bibnamefont {Schneeloch}}, \bibinfo
  {author} {\bibfnamefont {G.~D.}\ \bibnamefont {Gu}}, \bibinfo {author}
  {\bibfnamefont {F.}~\bibnamefont {Wang}},\ and\ \bibinfo {author}
  {\bibfnamefont {N.~L.}\ \bibnamefont {Wang}},\ }\bibfield  {title} {\bibinfo
  {title} {{Magnetoinfrared Spectroscopy of Landau Levels and Zeeman Splitting
  of Three-Dimensional Massless Dirac Fermions in ${\mathrm{ZrTe}}_{5}$}},\
  }\href {https://doi.org/10.1103/PhysRevLett.115.176404} {\bibfield  {journal}
  {\bibinfo  {journal} {Phys. Rev. Lett.}\ }\textbf {\bibinfo {volume} {115}},\
  \bibinfo {pages} {176404} (\bibinfo {year} {2015}{\natexlab{a}})}\BibitemShut
  {NoStop}%
\bibitem [{\citenamefont {Chen}\ \emph
  {et~al.}(2015{\natexlab{b}})\citenamefont {Chen}, \citenamefont {Zhang},
  \citenamefont {Schneeloch}, \citenamefont {Zhang}, \citenamefont {Li},
  \citenamefont {Gu},\ and\ \citenamefont
  {Wang}}]{dirac_chen-3DdiracOptical_oss3ddsz_2015}%
  \BibitemOpen
  \bibfield  {author} {\bibinfo {author} {\bibfnamefont {R.~Y.}\ \bibnamefont
  {Chen}}, \bibinfo {author} {\bibfnamefont {S.~J.}\ \bibnamefont {Zhang}},
  \bibinfo {author} {\bibfnamefont {J.~A.}\ \bibnamefont {Schneeloch}},
  \bibinfo {author} {\bibfnamefont {C.}~\bibnamefont {Zhang}}, \bibinfo
  {author} {\bibfnamefont {Q.}~\bibnamefont {Li}}, \bibinfo {author}
  {\bibfnamefont {G.~D.}\ \bibnamefont {Gu}},\ and\ \bibinfo {author}
  {\bibfnamefont {N.~L.}\ \bibnamefont {Wang}},\ }\bibfield  {title} {\bibinfo
  {title} {{Optical spectroscopy study of the three-dimensional Dirac semimetal
  ${\mathrm{ZrTe}}_{5}$}},\ }\href {https://doi.org/10.1103/PhysRevB.92.075107}
  {\bibfield  {journal} {\bibinfo  {journal} {Phys. Rev. B}\ }\textbf {\bibinfo
  {volume} {92}},\ \bibinfo {pages} {075107} (\bibinfo {year}
  {2015}{\natexlab{b}})}\BibitemShut {NoStop}%
\bibitem [{\citenamefont {Yuan}\ \emph {et~al.}(2016)\citenamefont {Yuan},
  \citenamefont {Zhang}, \citenamefont {Liu}, \citenamefont {Chaoyu},
  \citenamefont {Shen}, \citenamefont {Sui}, \citenamefont {Xu}, \citenamefont
  {Yu}, \citenamefont {An}, \citenamefont {Zhao}, \citenamefont {Yan},\ and\
  \citenamefont {Xiu}}]{dirac_xiang-diracMagOpt-oq2ddfz_2015}%
  \BibitemOpen
  \bibfield  {author} {\bibinfo {author} {\bibfnamefont {X.}~\bibnamefont
  {Yuan}}, \bibinfo {author} {\bibfnamefont {C.}~\bibnamefont {Zhang}},
  \bibinfo {author} {\bibfnamefont {Y.}~\bibnamefont {Liu}}, \bibinfo {author}
  {\bibfnamefont {S.}~\bibnamefont {Chaoyu}}, \bibinfo {author} {\bibfnamefont
  {S.}~\bibnamefont {Shen}}, \bibinfo {author} {\bibfnamefont {X.}~\bibnamefont
  {Sui}}, \bibinfo {author} {\bibfnamefont {J.}~\bibnamefont {Xu}}, \bibinfo
  {author} {\bibfnamefont {H.}~\bibnamefont {Yu}}, \bibinfo {author}
  {\bibfnamefont {Z.}~\bibnamefont {An}}, \bibinfo {author} {\bibfnamefont
  {J.}~\bibnamefont {Zhao}}, \bibinfo {author} {\bibfnamefont {S.}~\bibnamefont
  {Yan}},\ and\ \bibinfo {author} {\bibfnamefont {F.}~\bibnamefont {Xiu}},\
  }\bibfield  {title} {\bibinfo {title} {{Observation of quasi-two-dimensional
  Dirac fermions in ZrTe$_5$}},\ }\href {https://doi.org/10.1038/am.2016.166}
  {\bibfield  {journal} {\bibinfo  {journal} {NPG Asia Materials}\ }\textbf
  {\bibinfo {volume} {8}},\ \bibinfo {pages} {e325} (\bibinfo {year}
  {2016})}\BibitemShut {NoStop}%
\bibitem [{\citenamefont {Chen}\ \emph {et~al.}(2017)\citenamefont {Chen},
  \citenamefont {Chen}, \citenamefont {Zhong}, \citenamefont {Schneeloch},
  \citenamefont {Zhang}, \citenamefont {Huang}, \citenamefont {Qu},
  \citenamefont {Yu}, \citenamefont {Li}, \citenamefont {Gu},\ and\
  \citenamefont {Wang}}]{dirac-and-wti_chen-3DDiracIR-sebbi3dmd_2017}%
  \BibitemOpen
  \bibfield  {author} {\bibinfo {author} {\bibfnamefont {Z.-G.}\ \bibnamefont
  {Chen}}, \bibinfo {author} {\bibfnamefont {R.~Y.}\ \bibnamefont {Chen}},
  \bibinfo {author} {\bibfnamefont {R.~D.}\ \bibnamefont {Zhong}}, \bibinfo
  {author} {\bibfnamefont {J.}~\bibnamefont {Schneeloch}}, \bibinfo {author}
  {\bibfnamefont {C.}~\bibnamefont {Zhang}}, \bibinfo {author} {\bibfnamefont
  {Y.}~\bibnamefont {Huang}}, \bibinfo {author} {\bibfnamefont
  {F.}~\bibnamefont {Qu}}, \bibinfo {author} {\bibfnamefont {R.}~\bibnamefont
  {Yu}}, \bibinfo {author} {\bibfnamefont {Q.}~\bibnamefont {Li}}, \bibinfo
  {author} {\bibfnamefont {G.~D.}\ \bibnamefont {Gu}},\ and\ \bibinfo {author}
  {\bibfnamefont {N.~L.}\ \bibnamefont {Wang}},\ }\bibfield  {title} {\bibinfo
  {title} {{Spectroscopic evidence for bulk-band inversion and
  three-dimensional massive Dirac fermions in ZrTe$_5$}},\ }\href
  {https://doi.org/10.1073/pnas.1613110114} {\bibfield  {journal} {\bibinfo
  {journal} {Proceedings of the National Academy of Sciences}\ }\textbf
  {\bibinfo {volume} {114}},\ \bibinfo {pages} {816} (\bibinfo {year}
  {2017})}\BibitemShut {NoStop}%
\bibitem [{\citenamefont {Li}\ \emph {et~al.}(2016{\natexlab{a}})\citenamefont
  {Li}, \citenamefont {Kharzeev}, \citenamefont {Zhang}, \citenamefont {Huang},
  \citenamefont {Pletikosić}, \citenamefont {Fedorov}, \citenamefont {Zhong},
  \citenamefont {Schneeloch}, \citenamefont {Gu},\ and\ \citenamefont
  {Valla}}]{dirac-chiral_qiang-chiarMagnetic-cmez_2016}%
  \BibitemOpen
  \bibfield  {author} {\bibinfo {author} {\bibfnamefont {Q.}~\bibnamefont
  {Li}}, \bibinfo {author} {\bibfnamefont {D.}~\bibnamefont {Kharzeev}},
  \bibinfo {author} {\bibfnamefont {C.}~\bibnamefont {Zhang}}, \bibinfo
  {author} {\bibfnamefont {Y.}~\bibnamefont {Huang}}, \bibinfo {author}
  {\bibfnamefont {I.}~\bibnamefont {Pletikosić}}, \bibinfo {author}
  {\bibfnamefont {A.}~\bibnamefont {Fedorov}}, \bibinfo {author} {\bibfnamefont
  {R.}~\bibnamefont {Zhong}}, \bibinfo {author} {\bibfnamefont
  {J.}~\bibnamefont {Schneeloch}}, \bibinfo {author} {\bibfnamefont
  {G.}~\bibnamefont {Gu}},\ and\ \bibinfo {author} {\bibfnamefont
  {T.}~\bibnamefont {Valla}},\ }\bibfield  {title} {\bibinfo {title} {{Chiral
  magnetic effect in ZrTe$_5$}},\ }\href {https://doi.org/10.1038/nphys3648}
  {\bibfield  {journal} {\bibinfo  {journal} {Nature Physics}\ }\textbf
  {\bibinfo {volume} {12}},\ \bibinfo {pages} {550} (\bibinfo {year}
  {2016}{\natexlab{a}})}\BibitemShut {NoStop}%
\bibitem [{\citenamefont {Zheng}\ \emph {et~al.}(2016)\citenamefont {Zheng},
  \citenamefont {Lu}, \citenamefont {Zhu}, \citenamefont {Ning}, \citenamefont
  {Han}, \citenamefont {Zhang}, \citenamefont {Zhang}, \citenamefont {Xi},
  \citenamefont {Yang}, \citenamefont {Du}, \citenamefont {Yang}, \citenamefont
  {Zhang},\ and\ \citenamefont
  {Tian}}]{dirac-chiral_guolin-diracMagNeg-te3ddspz_2016}%
  \BibitemOpen
  \bibfield  {author} {\bibinfo {author} {\bibfnamefont {G.}~\bibnamefont
  {Zheng}}, \bibinfo {author} {\bibfnamefont {J.}~\bibnamefont {Lu}}, \bibinfo
  {author} {\bibfnamefont {X.}~\bibnamefont {Zhu}}, \bibinfo {author}
  {\bibfnamefont {W.}~\bibnamefont {Ning}}, \bibinfo {author} {\bibfnamefont
  {Y.}~\bibnamefont {Han}}, \bibinfo {author} {\bibfnamefont {H.}~\bibnamefont
  {Zhang}}, \bibinfo {author} {\bibfnamefont {J.}~\bibnamefont {Zhang}},
  \bibinfo {author} {\bibfnamefont {C.}~\bibnamefont {Xi}}, \bibinfo {author}
  {\bibfnamefont {J.}~\bibnamefont {Yang}}, \bibinfo {author} {\bibfnamefont
  {H.}~\bibnamefont {Du}}, \bibinfo {author} {\bibfnamefont {K.}~\bibnamefont
  {Yang}}, \bibinfo {author} {\bibfnamefont {Y.}~\bibnamefont {Zhang}},\ and\
  \bibinfo {author} {\bibfnamefont {M.}~\bibnamefont {Tian}},\ }\bibfield
  {title} {\bibinfo {title} {Transport evidence for the three-dimensional dirac
  semimetal phase in $\mathrm{ZrT}{\mathrm{e}}_{5}$},\ }\href
  {https://doi.org/10.1103/PhysRevB.93.115414} {\bibfield  {journal} {\bibinfo
  {journal} {Phys. Rev. B}\ }\textbf {\bibinfo {volume} {93}},\ \bibinfo
  {pages} {115414} (\bibinfo {year} {2016})}\BibitemShut {NoStop}%
\bibitem [{\citenamefont {Moreschini}\ \emph {et~al.}(2016)\citenamefont
  {Moreschini}, \citenamefont {Johannsen}, \citenamefont {Berger},
  \citenamefont {Denlinger}, \citenamefont {Jozwiak}, \citenamefont
  {Rotenberg}, \citenamefont {Kim}, \citenamefont {Bostwick},\ and\
  \citenamefont {Grioni}}]{arpes_moreschini-arpesWTI-ntlesz_2016}%
  \BibitemOpen
  \bibfield  {author} {\bibinfo {author} {\bibfnamefont {L.}~\bibnamefont
  {Moreschini}}, \bibinfo {author} {\bibfnamefont {J.~C.}\ \bibnamefont
  {Johannsen}}, \bibinfo {author} {\bibfnamefont {H.}~\bibnamefont {Berger}},
  \bibinfo {author} {\bibfnamefont {J.}~\bibnamefont {Denlinger}}, \bibinfo
  {author} {\bibfnamefont {C.}~\bibnamefont {Jozwiak}}, \bibinfo {author}
  {\bibfnamefont {E.}~\bibnamefont {Rotenberg}}, \bibinfo {author}
  {\bibfnamefont {K.~S.}\ \bibnamefont {Kim}}, \bibinfo {author} {\bibfnamefont
  {A.}~\bibnamefont {Bostwick}},\ and\ \bibinfo {author} {\bibfnamefont
  {M.}~\bibnamefont {Grioni}},\ }\bibfield  {title} {\bibinfo {title} {{Nature
  and topology of the low-energy states in ${\mathrm{ZrTe}}_{5}$}},\ }\href
  {https://doi.org/10.1103/PhysRevB.94.081101} {\bibfield  {journal} {\bibinfo
  {journal} {Phys. Rev. B}\ }\textbf {\bibinfo {volume} {94}},\ \bibinfo
  {pages} {081101(R)} (\bibinfo {year} {2016})}\BibitemShut {NoStop}%
\bibitem [{\citenamefont {Manzoni}\ \emph {et~al.}(2016)\citenamefont
  {Manzoni}, \citenamefont {Gragnaniello}, \citenamefont {Aut\`es},
  \citenamefont {Kuhn}, \citenamefont {Sterzi}, \citenamefont {Cilento},
  \citenamefont {Zacchigna}, \citenamefont {Enenkel}, \citenamefont {Vobornik},
  \citenamefont {Barba}, \citenamefont {Bisti}, \citenamefont {Bugnon},
  \citenamefont {Magrez}, \citenamefont {Strocov}, \citenamefont {Berger},
  \citenamefont {Yazyev}, \citenamefont {Fonin}, \citenamefont {Parmigiani},\
  and\ \citenamefont {Crepaldi}}]{bandstruct_manzoni-STIarpes-estipz_2016}%
  \BibitemOpen
  \bibfield  {author} {\bibinfo {author} {\bibfnamefont {G.}~\bibnamefont
  {Manzoni}}, \bibinfo {author} {\bibfnamefont {L.}~\bibnamefont
  {Gragnaniello}}, \bibinfo {author} {\bibfnamefont {G.}~\bibnamefont
  {Aut\`es}}, \bibinfo {author} {\bibfnamefont {T.}~\bibnamefont {Kuhn}},
  \bibinfo {author} {\bibfnamefont {A.}~\bibnamefont {Sterzi}}, \bibinfo
  {author} {\bibfnamefont {F.}~\bibnamefont {Cilento}}, \bibinfo {author}
  {\bibfnamefont {M.}~\bibnamefont {Zacchigna}}, \bibinfo {author}
  {\bibfnamefont {V.}~\bibnamefont {Enenkel}}, \bibinfo {author} {\bibfnamefont
  {I.}~\bibnamefont {Vobornik}}, \bibinfo {author} {\bibfnamefont
  {L.}~\bibnamefont {Barba}}, \bibinfo {author} {\bibfnamefont
  {F.}~\bibnamefont {Bisti}}, \bibinfo {author} {\bibfnamefont
  {P.}~\bibnamefont {Bugnon}}, \bibinfo {author} {\bibfnamefont
  {A.}~\bibnamefont {Magrez}}, \bibinfo {author} {\bibfnamefont {V.~N.}\
  \bibnamefont {Strocov}}, \bibinfo {author} {\bibfnamefont {H.}~\bibnamefont
  {Berger}}, \bibinfo {author} {\bibfnamefont {O.~V.}\ \bibnamefont {Yazyev}},
  \bibinfo {author} {\bibfnamefont {M.}~\bibnamefont {Fonin}}, \bibinfo
  {author} {\bibfnamefont {F.}~\bibnamefont {Parmigiani}},\ and\ \bibinfo
  {author} {\bibfnamefont {A.}~\bibnamefont {Crepaldi}},\ }\bibfield  {title}
  {\bibinfo {title} {{Evidence for a Strong Topological Insulator Phase in
  ${\mathrm{ZrTe}}_{5}$}},\ }\href
  {https://doi.org/10.1103/PhysRevLett.117.237601} {\bibfield  {journal}
  {\bibinfo  {journal} {Phys. Rev. Lett.}\ }\textbf {\bibinfo {volume} {117}},\
  \bibinfo {pages} {237601} (\bibinfo {year} {2016})}\BibitemShut {NoStop}%
\bibitem [{\citenamefont {Manzoni}\ \emph {et~al.}(2017)\citenamefont
  {Manzoni}, \citenamefont {Crepaldi}, \citenamefont {Autès}, \citenamefont
  {Sterzi}, \citenamefont {Cilento}, \citenamefont {Akrap}, \citenamefont
  {Vobornik}, \citenamefont {Gragnaniello}, \citenamefont {Bugnon},
  \citenamefont {Fonin}, \citenamefont {Berger}, \citenamefont {Zacchigna},
  \citenamefont {Yazyev},\ and\ \citenamefont
  {Parmigiani}}]{arpes_manzoni-ArpesSTI-tdnmbsz_2017}%
  \BibitemOpen
  \bibfield  {author} {\bibinfo {author} {\bibfnamefont {G.}~\bibnamefont
  {Manzoni}}, \bibinfo {author} {\bibfnamefont {A.}~\bibnamefont {Crepaldi}},
  \bibinfo {author} {\bibfnamefont {G.}~\bibnamefont {Autès}}, \bibinfo
  {author} {\bibfnamefont {A.}~\bibnamefont {Sterzi}}, \bibinfo {author}
  {\bibfnamefont {F.}~\bibnamefont {Cilento}}, \bibinfo {author} {\bibfnamefont
  {A.}~\bibnamefont {Akrap}}, \bibinfo {author} {\bibfnamefont
  {I.}~\bibnamefont {Vobornik}}, \bibinfo {author} {\bibfnamefont
  {L.}~\bibnamefont {Gragnaniello}}, \bibinfo {author} {\bibfnamefont
  {P.}~\bibnamefont {Bugnon}}, \bibinfo {author} {\bibfnamefont
  {M.}~\bibnamefont {Fonin}}, \bibinfo {author} {\bibfnamefont
  {H.}~\bibnamefont {Berger}}, \bibinfo {author} {\bibfnamefont
  {M.}~\bibnamefont {Zacchigna}}, \bibinfo {author} {\bibfnamefont
  {O.}~\bibnamefont {Yazyev}},\ and\ \bibinfo {author} {\bibfnamefont
  {F.}~\bibnamefont {Parmigiani}},\ }\bibfield  {title} {\bibinfo {title}
  {{Temperature dependent non-monotonic bands shift in ZrTe$_5$}},\ }\href
  {https://doi.org/10.1016/j.elspec.2016.09.006} {\bibfield  {journal}
  {\bibinfo  {journal} {Journal of Electron Spectroscopy and Related
  Phenomena}\ }\textbf {\bibinfo {volume} {219}},\ \bibinfo {pages} {9}
  (\bibinfo {year} {2017})}\BibitemShut {NoStop}%
\bibitem [{\citenamefont {Zhang}\ \emph
  {et~al.}(2017{\natexlab{a}})\citenamefont {Zhang}, \citenamefont {Wang},
  \citenamefont {Yu}, \citenamefont {Liu}, \citenamefont {Liang}, \citenamefont
  {Huang}, \citenamefont {Nie}, \citenamefont {Sun}, \citenamefont {Zhang},
  \citenamefont {Shen} \emph
  {et~al.}}]{bandstruct-and-fit_zhang-arpeslifshitz-eetilttnz_2017}%
  \BibitemOpen
  \bibfield  {author} {\bibinfo {author} {\bibfnamefont {Y.}~\bibnamefont
  {Zhang}}, \bibinfo {author} {\bibfnamefont {C.}~\bibnamefont {Wang}},
  \bibinfo {author} {\bibfnamefont {L.}~\bibnamefont {Yu}}, \bibinfo {author}
  {\bibfnamefont {G.}~\bibnamefont {Liu}}, \bibinfo {author} {\bibfnamefont
  {A.}~\bibnamefont {Liang}}, \bibinfo {author} {\bibfnamefont
  {J.}~\bibnamefont {Huang}}, \bibinfo {author} {\bibfnamefont
  {S.}~\bibnamefont {Nie}}, \bibinfo {author} {\bibfnamefont {X.}~\bibnamefont
  {Sun}}, \bibinfo {author} {\bibfnamefont {Y.}~\bibnamefont {Zhang}}, \bibinfo
  {author} {\bibfnamefont {B.}~\bibnamefont {Shen}}, \emph {et~al.},\
  }\bibfield  {title} {{\selectlanguage {english}\bibinfo {title} {{Electronic
  evidence of temperature-induced Lifshitz transition and topological nature in
  ZrTe$_5$}}},\ }\href {https://doi.org/10.1038/ncomms15512} {\bibfield
  {journal} {\bibinfo  {journal} {Nature Communications}\ }\textbf {\bibinfo
  {volume} {8}},\ \bibinfo {pages} {15512} (\bibinfo {year}
  {2017}{\natexlab{a}})}\BibitemShut {NoStop}%
\bibitem [{\citenamefont {Shahi}\ \emph {et~al.}(2018)\citenamefont {Shahi},
  \citenamefont {Singh}, \citenamefont {Sun}, \citenamefont {Zhao},
  \citenamefont {Chen}, \citenamefont {Lv}, \citenamefont {Li}, \citenamefont
  {Yan}, \citenamefont {Mandrus},\ and\ \citenamefont
  {Cheng}}]{fit_shahi-bipolar-bcpoetp_2018}%
  \BibitemOpen
  \bibfield  {author} {\bibinfo {author} {\bibfnamefont {P.}~\bibnamefont
  {Shahi}}, \bibinfo {author} {\bibfnamefont {D.~J.}\ \bibnamefont {Singh}},
  \bibinfo {author} {\bibfnamefont {J.~P.}\ \bibnamefont {Sun}}, \bibinfo
  {author} {\bibfnamefont {L.~X.}\ \bibnamefont {Zhao}}, \bibinfo {author}
  {\bibfnamefont {G.~F.}\ \bibnamefont {Chen}}, \bibinfo {author}
  {\bibfnamefont {Y.~Y.}\ \bibnamefont {Lv}}, \bibinfo {author} {\bibfnamefont
  {J.}~\bibnamefont {Li}}, \bibinfo {author} {\bibfnamefont {J.-Q.}\
  \bibnamefont {Yan}}, \bibinfo {author} {\bibfnamefont {D.~G.}\ \bibnamefont
  {Mandrus}},\ and\ \bibinfo {author} {\bibfnamefont {J.-G.}\ \bibnamefont
  {Cheng}},\ }\bibfield  {title} {\bibinfo {title} {{Bipolar Conduction as the
  Possible Origin of the Electronic Transition in Pentatellurides: Metallic vs
  Semiconducting Behavior}},\ }\href
  {https://doi.org/10.1103/PhysRevX.8.021055} {\bibfield  {journal} {\bibinfo
  {journal} {Phys. Rev. X}\ }\textbf {\bibinfo {volume} {8}},\ \bibinfo {pages}
  {021055} (\bibinfo {year} {2018})}\BibitemShut {NoStop}%
\bibitem [{\citenamefont {Fan}\ \emph {et~al.}(2017)\citenamefont {Fan},
  \citenamefont {Liang}, \citenamefont {Chen}, \citenamefont {Yao},\ and\
  \citenamefont {Zhou}}]{bandstruct_fan-WTIvSTI-tbswtizh_2017}%
  \BibitemOpen
  \bibfield  {author} {\bibinfo {author} {\bibfnamefont {Z.}~\bibnamefont
  {Fan}}, \bibinfo {author} {\bibfnamefont {Q.-F.}\ \bibnamefont {Liang}},
  \bibinfo {author} {\bibfnamefont {Y.~B.}\ \bibnamefont {Chen}}, \bibinfo
  {author} {\bibfnamefont {S.-H.}\ \bibnamefont {Yao}},\ and\ \bibinfo {author}
  {\bibfnamefont {J.}~\bibnamefont {Zhou}},\ }\bibfield  {title}
  {{\selectlanguage {english}\bibinfo {title} {{Transition between strong and weak
  topological insulator in ZrTe$_5$ and HfTe$_5$}}},\ }\href
  {https://doi.org/10.1038/srep45667} {\bibfield  {journal} {\bibinfo
  {journal} {Scientific Reports}\ }\textbf {\bibinfo {volume} {7}},\ \bibinfo
  {pages} {45667} (\bibinfo {year} {2017})}\BibitemShut {NoStop}%
\bibitem [{\citenamefont {Mutch}\ \emph {et~al.}(2019)\citenamefont {Mutch},
  \citenamefont {Chen}, \citenamefont {Went}, \citenamefont {Qian},
  \citenamefont {Wilson}, \citenamefont {Andreev}, \citenamefont {Chen},\ and\
  \citenamefont
  {Chu}}]{dirac-and-sti-and-wti_mutch-strainTunedTransition-esttptz_2019}%
  \BibitemOpen
  \bibfield  {author} {\bibinfo {author} {\bibfnamefont {J.}~\bibnamefont
  {Mutch}}, \bibinfo {author} {\bibfnamefont {W.-C.}\ \bibnamefont {Chen}},
  \bibinfo {author} {\bibfnamefont {P.}~\bibnamefont {Went}}, \bibinfo {author}
  {\bibfnamefont {T.}~\bibnamefont {Qian}}, \bibinfo {author} {\bibfnamefont
  {I.~Z.}\ \bibnamefont {Wilson}}, \bibinfo {author} {\bibfnamefont
  {A.}~\bibnamefont {Andreev}}, \bibinfo {author} {\bibfnamefont {C.-C.}\
  \bibnamefont {Chen}},\ and\ \bibinfo {author} {\bibfnamefont {J.-H.}\
  \bibnamefont {Chu}},\ }\bibfield  {title} {\bibinfo {title} {{Evidence for a
  strain-tuned topological phase transition in ZrTe$_5$}},\ }\href
  {https://doi.org/10.1126/sciadv.aav9771} {\bibfield  {journal} {\bibinfo
  {journal} {Science Advances}\ }\textbf {\bibinfo {volume} {5}},\ \bibinfo
  {pages} {eaav9771} (\bibinfo {year} {2019})}\BibitemShut {NoStop}%
\bibitem [{\citenamefont {Wang}\ \emph {et~al.}(2022)\citenamefont {Wang},
  \citenamefont {Legg}, \citenamefont {B\"omerich}, \citenamefont {Park},
  \citenamefont {Biesenkamp}, \citenamefont {Taskin}, \citenamefont {Braden},
  \citenamefont {Rosch},\ and\ \citenamefont
  {Ando}}]{sdho-torus-flux_wang-magchiralaniso-gmatsz_2022}%
  \BibitemOpen
  \bibfield  {author} {\bibinfo {author} {\bibfnamefont {Y.}~\bibnamefont
  {Wang}}, \bibinfo {author} {\bibfnamefont {H.~F.}\ \bibnamefont {Legg}},
  \bibinfo {author} {\bibfnamefont {T.}~\bibnamefont {B\"omerich}}, \bibinfo
  {author} {\bibfnamefont {J.}~\bibnamefont {Park}}, \bibinfo {author}
  {\bibfnamefont {S.}~\bibnamefont {Biesenkamp}}, \bibinfo {author}
  {\bibfnamefont {A.~A.}\ \bibnamefont {Taskin}}, \bibinfo {author}
  {\bibfnamefont {M.}~\bibnamefont {Braden}}, \bibinfo {author} {\bibfnamefont
  {A.}~\bibnamefont {Rosch}},\ and\ \bibinfo {author} {\bibfnamefont
  {Y.}~\bibnamefont {Ando}},\ }\bibfield  {title} {\bibinfo {title} {{Gigantic
  Magnetochiral Anisotropy in the Topological Semimetal
  ${\mathrm{ZrTe}}_{5}$}},\ }\href
  {https://doi.org/10.1103/PhysRevLett.128.176602} {\bibfield  {journal}
  {\bibinfo  {journal} {Phys. Rev. Lett.}\ }\textbf {\bibinfo {volume} {128}},\
  \bibinfo {pages} {176602} (\bibinfo {year} {2022})}\BibitemShut {NoStop}%
\bibitem [{\citenamefont {Weng}\ \emph {et~al.}(2014)\citenamefont {Weng},
  \citenamefont {Dai},\ and\ \citenamefont
  {Fang}}]{bandstruct_weng-QSHparadigm-tmpzh_2014}%
  \BibitemOpen
  \bibfield  {author} {\bibinfo {author} {\bibfnamefont {H.}~\bibnamefont
  {Weng}}, \bibinfo {author} {\bibfnamefont {X.}~\bibnamefont {Dai}},\ and\
  \bibinfo {author} {\bibfnamefont {Z.}~\bibnamefont {Fang}},\ }\bibfield
  {title} {\bibinfo {title} {Transition-metal pentatelluride
  $\mathrm{ZrTe}{}_{5}$ and $\mathrm{HfTe}{}_{5}$: A paradigm for large-gap
  quantum spin hall insulators},\ }\href
  {https://doi.org/10.1103/PhysRevX.4.011002} {\bibfield  {journal} {\bibinfo
  {journal} {Phys. Rev. X}\ }\textbf {\bibinfo {volume} {4}},\ \bibinfo {pages}
  {011002} (\bibinfo {year} {2014})}\BibitemShut {NoStop}%
\bibitem [{\citenamefont {Fjellv{\aa}g}\ and\ \citenamefont
  {Kjekshus}(1986)}]{lattice_fjellvag-powDifExp-spzhpd_1986}%
  \BibitemOpen
  \bibfield  {author} {\bibinfo {author} {\bibfnamefont {H.}~\bibnamefont
  {Fjellv{\aa}g}}\ and\ \bibinfo {author} {\bibfnamefont {A.}~\bibnamefont
  {Kjekshus}},\ }\bibfield  {title} {\bibinfo {title} {{Structural properties
  of ZrTe$_5$ and HfTe$_5$ as seen by powder diffraction}},\ }\href
  {https://doi.org/10.1016/0038-1098(86)90536-3} {\bibfield  {journal}
  {\bibinfo  {journal} {Solid State Communications}\ }\textbf {\bibinfo
  {volume} {60}},\ \bibinfo {pages} {91} (\bibinfo {year} {1986})}\BibitemShut
  {NoStop}%
\bibitem [{\citenamefont {Konstantinova}\ \emph {et~al.}(2020)\citenamefont
  {Konstantinova}, \citenamefont {Wu}, \citenamefont {Yin}, \citenamefont
  {Tao}, \citenamefont {Gu}, \citenamefont {Wang}, \citenamefont {Yang},
  \citenamefont {Zaliznyak},\ and\ \citenamefont
  {Zhu}}]{bandstruct-dirac_sun-pumpProbeDirac-pdsz_2020}%
  \BibitemOpen
  \bibfield  {author} {\bibinfo {author} {\bibfnamefont {T.}~\bibnamefont
  {Konstantinova}}, \bibinfo {author} {\bibfnamefont {L.}~\bibnamefont {Wu}},
  \bibinfo {author} {\bibfnamefont {W.-G.}\ \bibnamefont {Yin}}, \bibinfo
  {author} {\bibfnamefont {J.}~\bibnamefont {Tao}}, \bibinfo {author}
  {\bibfnamefont {G.~D.}\ \bibnamefont {Gu}}, \bibinfo {author} {\bibfnamefont
  {X.~J.}\ \bibnamefont {Wang}}, \bibinfo {author} {\bibfnamefont
  {J.}~\bibnamefont {Yang}}, \bibinfo {author} {\bibfnamefont {I.~A.}\
  \bibnamefont {Zaliznyak}},\ and\ \bibinfo {author} {\bibfnamefont
  {Y.}~\bibnamefont {Zhu}},\ }\bibfield  {title} {\bibinfo {title}
  {{Photoinduced Dirac semimetal in ZrTe$_5$}},\ }\href
  {https://doi.org/10.1038/s41535-020-00280-8} {\bibfield  {journal} {\bibinfo
  {journal} {npj Quantum Materials}\ }\textbf {\bibinfo {volume} {5}},\
  \bibinfo {pages} {80} (\bibinfo {year} {2020})}\BibitemShut {NoStop}%
\bibitem [{\citenamefont {Wu}\ \emph {et~al.}(2016)\citenamefont {Wu},
  \citenamefont {Ma}, \citenamefont {Nie}, \citenamefont {Zhao}, \citenamefont
  {Huang}, \citenamefont {Yin}, \citenamefont {Fu}, \citenamefont {Richard},
  \citenamefont {Chen}, \citenamefont {Fang}, \citenamefont {Dai},
  \citenamefont {Weng}, \citenamefont {Qian}, \citenamefont {Ding},\ and\
  \citenamefont {Pan}}]{stmEdge_we-stmWTI-etesleg_2016}%
  \BibitemOpen
  \bibfield  {author} {\bibinfo {author} {\bibfnamefont {R.}~\bibnamefont
  {Wu}}, \bibinfo {author} {\bibfnamefont {J.-Z.}\ \bibnamefont {Ma}}, \bibinfo
  {author} {\bibfnamefont {S.-M.}\ \bibnamefont {Nie}}, \bibinfo {author}
  {\bibfnamefont {L.-X.}\ \bibnamefont {Zhao}}, \bibinfo {author}
  {\bibfnamefont {X.}~\bibnamefont {Huang}}, \bibinfo {author} {\bibfnamefont
  {J.-X.}\ \bibnamefont {Yin}}, \bibinfo {author} {\bibfnamefont {B.-B.}\
  \bibnamefont {Fu}}, \bibinfo {author} {\bibfnamefont {P.}~\bibnamefont
  {Richard}}, \bibinfo {author} {\bibfnamefont {G.-F.}\ \bibnamefont {Chen}},
  \bibinfo {author} {\bibfnamefont {Z.}~\bibnamefont {Fang}}, \bibinfo {author}
  {\bibfnamefont {X.}~\bibnamefont {Dai}}, \bibinfo {author} {\bibfnamefont
  {H.-M.}\ \bibnamefont {Weng}}, \bibinfo {author} {\bibfnamefont
  {T.}~\bibnamefont {Qian}}, \bibinfo {author} {\bibfnamefont {H.}~\bibnamefont
  {Ding}},\ and\ \bibinfo {author} {\bibfnamefont {S.~H.}\ \bibnamefont
  {Pan}},\ }\bibfield  {title} {\bibinfo {title} {{Evidence for Topological
  Edge States in a Large Energy Gap near the Step Edges on the Surface of
  ${\mathrm{ZrTe}}_{5}$}},\ }\href {https://doi.org/10.1103/PhysRevX.6.021017}
  {\bibfield  {journal} {\bibinfo  {journal} {Phys. Rev. X}\ }\textbf {\bibinfo
  {volume} {6}},\ \bibinfo {pages} {021017} (\bibinfo {year}
  {2016})}\BibitemShut {NoStop}%
\bibitem [{\citenamefont {Li}\ \emph {et~al.}(2016{\natexlab{b}})\citenamefont
  {Li}, \citenamefont {Huang}, \citenamefont {Lv}, \citenamefont {Zhang},
  \citenamefont {Yang}, \citenamefont {Zhang}, \citenamefont {Chen},
  \citenamefont {Yao}, \citenamefont {Zhou}, \citenamefont {Lu}, \citenamefont
  {Sheng}, \citenamefont {Li}, \citenamefont {Jia}, \citenamefont {Xue},
  \citenamefont {Chen},\ and\ \citenamefont
  {Xing}}]{stmEdge_xbing-stmWTI-eoteds_2016}%
  \BibitemOpen
  \bibfield  {author} {\bibinfo {author} {\bibfnamefont {X.-B.}\ \bibnamefont
  {Li}}, \bibinfo {author} {\bibfnamefont {W.-K.}\ \bibnamefont {Huang}},
  \bibinfo {author} {\bibfnamefont {Y.-Y.}\ \bibnamefont {Lv}}, \bibinfo
  {author} {\bibfnamefont {K.-W.}\ \bibnamefont {Zhang}}, \bibinfo {author}
  {\bibfnamefont {C.-L.}\ \bibnamefont {Yang}}, \bibinfo {author}
  {\bibfnamefont {B.-B.}\ \bibnamefont {Zhang}}, \bibinfo {author}
  {\bibfnamefont {Y.~B.}\ \bibnamefont {Chen}}, \bibinfo {author}
  {\bibfnamefont {S.-H.}\ \bibnamefont {Yao}}, \bibinfo {author} {\bibfnamefont
  {J.}~\bibnamefont {Zhou}}, \bibinfo {author} {\bibfnamefont {M.-H.}\
  \bibnamefont {Lu}}, \bibinfo {author} {\bibfnamefont {L.}~\bibnamefont
  {Sheng}}, \bibinfo {author} {\bibfnamefont {S.-C.}\ \bibnamefont {Li}},
  \bibinfo {author} {\bibfnamefont {J.-F.}\ \bibnamefont {Jia}}, \bibinfo
  {author} {\bibfnamefont {Q.-K.}\ \bibnamefont {Xue}}, \bibinfo {author}
  {\bibfnamefont {Y.-F.}\ \bibnamefont {Chen}},\ and\ \bibinfo {author}
  {\bibfnamefont {D.-Y.}\ \bibnamefont {Xing}},\ }\bibfield  {title} {\bibinfo
  {title} {{Experimental Observation of Topological Edge States at the Surface
  Step Edge of the Topological Insulator ${\mathrm{ZrTe}}_{5}$}},\ }\href
  {https://doi.org/10.1103/PhysRevLett.116.176803} {\bibfield  {journal}
  {\bibinfo  {journal} {Phys. Rev. Lett.}\ }\textbf {\bibinfo {volume} {116}},\
  \bibinfo {pages} {176803} (\bibinfo {year} {2016}{\natexlab{b}})}\BibitemShut
  {NoStop}%
\bibitem [{\citenamefont {Liu}\ \emph {et~al.}(2016)\citenamefont {Liu},
  \citenamefont {Yuan}, \citenamefont {Zhang}, \citenamefont {Jin},
  \citenamefont {Narayan}, \citenamefont {Luo}, \citenamefont {Chen},
  \citenamefont {Yang}, \citenamefont {Zou}, \citenamefont {Wu}, \citenamefont
  {Sanvito}, \citenamefont {Xia}, \citenamefont {Li}, \citenamefont {Wang},\
  and\ \citenamefont {Xiu}}]{bandstruct-and-fit_liu-dynmass-zsdmgdsz_2016}%
  \BibitemOpen
  \bibfield  {author} {\bibinfo {author} {\bibfnamefont {Y.}~\bibnamefont
  {Liu}}, \bibinfo {author} {\bibfnamefont {X.}~\bibnamefont {Yuan}}, \bibinfo
  {author} {\bibfnamefont {C.}~\bibnamefont {Zhang}}, \bibinfo {author}
  {\bibfnamefont {Z.}~\bibnamefont {Jin}}, \bibinfo {author} {\bibfnamefont
  {A.}~\bibnamefont {Narayan}}, \bibinfo {author} {\bibfnamefont
  {C.}~\bibnamefont {Luo}}, \bibinfo {author} {\bibfnamefont {Z.}~\bibnamefont
  {Chen}}, \bibinfo {author} {\bibfnamefont {L.}~\bibnamefont {Yang}}, \bibinfo
  {author} {\bibfnamefont {J.}~\bibnamefont {Zou}}, \bibinfo {author}
  {\bibfnamefont {X.}~\bibnamefont {Wu}}, \bibinfo {author} {\bibfnamefont
  {S.}~\bibnamefont {Sanvito}}, \bibinfo {author} {\bibfnamefont
  {Z.}~\bibnamefont {Xia}}, \bibinfo {author} {\bibfnamefont {L.}~\bibnamefont
  {Li}}, \bibinfo {author} {\bibfnamefont {Z.}~\bibnamefont {Wang}},\ and\
  \bibinfo {author} {\bibfnamefont {F.}~\bibnamefont {Xiu}},\ }\bibfield
  {title} {{\selectlanguage {english}\bibinfo {title} {{Zeeman splitting and
  dynamical mass generation in Dirac semimetal ZrTe$_5$}}},\ }\href
  {https://doi.org/10.1038/ncomms12516} {\bibfield  {journal} {\bibinfo
  {journal} {Nature Communications}\ }\textbf {\bibinfo {volume} {7}},\
  \bibinfo {pages} {12516} (\bibinfo {year} {2016})}\BibitemShut {NoStop}%
\bibitem [{\citenamefont {Tian}\ \emph {et~al.}(2019)\citenamefont {Tian},
  \citenamefont {Ghassemi},\ and\ \citenamefont
  {Ross}}]{wti-sti_tian-bandinversion-debetbi_2019}%
  \BibitemOpen
  \bibfield  {author} {\bibinfo {author} {\bibfnamefont {Y.}~\bibnamefont
  {Tian}}, \bibinfo {author} {\bibfnamefont {N.}~\bibnamefont {Ghassemi}},\
  and\ \bibinfo {author} {\bibfnamefont {J.~H.}\ \bibnamefont {Ross}},\
  }\bibfield  {title} {\bibinfo {title} {{Dirac electron behavior and NMR
  evidence for topological band inversion in ${\mathrm{ZrTe}}_{5}$}},\ }\href
  {https://doi.org/10.1103/PhysRevB.100.165149} {\bibfield  {journal} {\bibinfo
   {journal} {Phys. Rev. B}\ }\textbf {\bibinfo {volume} {100}},\ \bibinfo
  {pages} {165149} (\bibinfo {year} {2019})}\BibitemShut {NoStop}%
\bibitem [{\citenamefont {Aryal}\ \emph {et~al.}(2021)\citenamefont {Aryal},
  \citenamefont {Jin}, \citenamefont {Li}, \citenamefont {Tsvelik},\ and\
  \citenamefont {Yin}}]{bandstruct-WTIvSTI-tptpsdts_2021}%
  \BibitemOpen
  \bibfield  {author} {\bibinfo {author} {\bibfnamefont {N.}~\bibnamefont
  {Aryal}}, \bibinfo {author} {\bibfnamefont {X.}~\bibnamefont {Jin}}, \bibinfo
  {author} {\bibfnamefont {Q.}~\bibnamefont {Li}}, \bibinfo {author}
  {\bibfnamefont {A.~M.}\ \bibnamefont {Tsvelik}},\ and\ \bibinfo {author}
  {\bibfnamefont {W.}~\bibnamefont {Yin}},\ }\bibfield  {title} {\bibinfo
  {title} {{Topological Phase Transition and Phonon-Space Dirac Topology
  Surfaces in ZrTe$_5$}},\ }\href
  {https://doi.org/10.1103/PhysRevLett.126.016401} {\bibfield  {journal}
  {\bibinfo  {journal} {Phys. Rev. Lett.}\ }\textbf {\bibinfo {volume} {126}},\
  \bibinfo {pages} {016401} (\bibinfo {year} {2021})}\BibitemShut {NoStop}%
\bibitem [{\citenamefont {Sun}\ \emph {et~al.}(2020)\citenamefont {Sun},
  \citenamefont {Cao}, \citenamefont {Cui}, \citenamefont {Zhu}, \citenamefont
  {Ma}, \citenamefont {Wang}, \citenamefont {Zhuo}, \citenamefont {Cheng},
  \citenamefont {Wang}, \citenamefont {Wan},\ and\ \citenamefont
  {Chen}}]{bandstruct-and-sdho-pressure_sun-AHEzeeman-lzsiahez_2020}%
  \BibitemOpen
  \bibfield  {author} {\bibinfo {author} {\bibfnamefont {Z.}~\bibnamefont
  {Sun}}, \bibinfo {author} {\bibfnamefont {Z.}~\bibnamefont {Cao}}, \bibinfo
  {author} {\bibfnamefont {J.}~\bibnamefont {Cui}}, \bibinfo {author}
  {\bibfnamefont {C.}~\bibnamefont {Zhu}}, \bibinfo {author} {\bibfnamefont
  {D.}~\bibnamefont {Ma}}, \bibinfo {author} {\bibfnamefont {H.}~\bibnamefont
  {Wang}}, \bibinfo {author} {\bibfnamefont {W.}~\bibnamefont {Zhuo}}, \bibinfo
  {author} {\bibfnamefont {Z.}~\bibnamefont {Cheng}}, \bibinfo {author}
  {\bibfnamefont {Z.}~\bibnamefont {Wang}}, \bibinfo {author} {\bibfnamefont
  {X.}~\bibnamefont {Wan}},\ and\ \bibinfo {author} {\bibfnamefont
  {X.}~\bibnamefont {Chen}},\ }\bibfield  {title} {\bibinfo {title} {{Large
  Zeeman splitting induced anomalous Hall effect in ZrTe$_5$}},\ }\href
  {https://doi.org/10.1038/s41535-020-0239-z} {\bibfield  {journal} {\bibinfo
  {journal} {npj Quantum Materials}\ }\textbf {\bibinfo {volume} {5}},\
  \bibinfo {pages} {36} (\bibinfo {year} {2020})}\BibitemShut {NoStop}%
\bibitem [{\citenamefont {Ge}\ \emph {et~al.}(2020)\citenamefont {Ge},
  \citenamefont {Ma}, \citenamefont {Liu}, \citenamefont {Wang}, \citenamefont
  {Li}, \citenamefont {Luo}, \citenamefont {Luo}, \citenamefont {Xing},
  \citenamefont {Yan}, \citenamefont {Mandrus}, \citenamefont {Liu},
  \citenamefont {Xie},\ and\ \citenamefont {Wang}}]{berry-flux_ju-uheibc_2020}%
  \BibitemOpen
  \bibfield  {author} {\bibinfo {author} {\bibfnamefont {J.}~\bibnamefont
  {Ge}}, \bibinfo {author} {\bibfnamefont {D.}~\bibnamefont {Ma}}, \bibinfo
  {author} {\bibfnamefont {Y.}~\bibnamefont {Liu}}, \bibinfo {author}
  {\bibfnamefont {H.}~\bibnamefont {Wang}}, \bibinfo {author} {\bibfnamefont
  {Y.}~\bibnamefont {Li}}, \bibinfo {author} {\bibfnamefont {J.}~\bibnamefont
  {Luo}}, \bibinfo {author} {\bibfnamefont {T.}~\bibnamefont {Luo}}, \bibinfo
  {author} {\bibfnamefont {Y.}~\bibnamefont {Xing}}, \bibinfo {author}
  {\bibfnamefont {J.}~\bibnamefont {Yan}}, \bibinfo {author} {\bibfnamefont
  {D.}~\bibnamefont {Mandrus}}, \bibinfo {author} {\bibfnamefont
  {H.}~\bibnamefont {Liu}}, \bibinfo {author} {\bibfnamefont {X.~C.}\
  \bibnamefont {Xie}},\ and\ \bibinfo {author} {\bibfnamefont {J.}~\bibnamefont
  {Wang}},\ }\bibfield  {title} {\bibinfo {title} {{Unconventional Hall effect
  induced by Berry curvature}},\ }\href {https://doi.org/10.1093/nsr/nwaa163}
  {\bibfield  {journal} {\bibinfo  {journal} {National Science Review}\
  }\textbf {\bibinfo {volume} {7}},\ \bibinfo {pages} {1879} (\bibinfo {year}
  {2020})}\BibitemShut {NoStop}%
\bibitem [{\citenamefont {Tang}\ \emph {et~al.}(2019)\citenamefont {Tang},
  \citenamefont {Ren}, \citenamefont {Wang}, \citenamefont {Zhong},
  \citenamefont {Schneeloch}, \citenamefont {Yang}, \citenamefont {Yang},
  \citenamefont {Lee}, \citenamefont {Gu}, \citenamefont {Qiao},\ and\
  \citenamefont {Zhang}}]{multiband_tang-3DQHE-3dqhemit_2019}%
  \BibitemOpen
  \bibfield  {author} {\bibinfo {author} {\bibfnamefont {F.}~\bibnamefont
  {Tang}}, \bibinfo {author} {\bibfnamefont {Y.}~\bibnamefont {Ren}}, \bibinfo
  {author} {\bibfnamefont {P.}~\bibnamefont {Wang}}, \bibinfo {author}
  {\bibfnamefont {R.}~\bibnamefont {Zhong}}, \bibinfo {author} {\bibfnamefont
  {J.}~\bibnamefont {Schneeloch}}, \bibinfo {author} {\bibfnamefont {S.~A.}\
  \bibnamefont {Yang}}, \bibinfo {author} {\bibfnamefont {K.}~\bibnamefont
  {Yang}}, \bibinfo {author} {\bibfnamefont {P.~A.}\ \bibnamefont {Lee}},
  \bibinfo {author} {\bibfnamefont {G.}~\bibnamefont {Gu}}, \bibinfo {author}
  {\bibfnamefont {Z.}~\bibnamefont {Qiao}},\ and\ \bibinfo {author}
  {\bibfnamefont {L.}~\bibnamefont {Zhang}},\ }\bibfield  {title} {\bibinfo
  {title} {{Three-dimensional quantum Hall effect and metal–insulator
  transition in ZrTe$_5$}},\ }\href {https://doi.org/10.1038/s41586-019-1180-9}
  {\bibfield  {journal} {\bibinfo  {journal} {Nature}\ }\textbf {\bibinfo
  {volume} {569}},\ \bibinfo {pages} {537–541} (\bibinfo {year}
  {2019})}\BibitemShut {NoStop}%
\bibitem [{\citenamefont {Lu}\ \emph {et~al.}(2017)\citenamefont {Lu},
  \citenamefont {Zheng}, \citenamefont {Zhu}, \citenamefont {Ning},
  \citenamefont {Zhang}, \citenamefont {Yang}, \citenamefont {Du},
  \citenamefont {Yang}, \citenamefont {Lu}, \citenamefont {Zhang},\ and\
  \citenamefont {Tian}}]{multiband-and-fit_lu-thicknessDep-tttbtns_2017}%
  \BibitemOpen
  \bibfield  {author} {\bibinfo {author} {\bibfnamefont {J.}~\bibnamefont
  {Lu}}, \bibinfo {author} {\bibfnamefont {G.}~\bibnamefont {Zheng}}, \bibinfo
  {author} {\bibfnamefont {X.}~\bibnamefont {Zhu}}, \bibinfo {author}
  {\bibfnamefont {W.}~\bibnamefont {Ning}}, \bibinfo {author} {\bibfnamefont
  {H.}~\bibnamefont {Zhang}}, \bibinfo {author} {\bibfnamefont
  {J.}~\bibnamefont {Yang}}, \bibinfo {author} {\bibfnamefont {H.}~\bibnamefont
  {Du}}, \bibinfo {author} {\bibfnamefont {K.}~\bibnamefont {Yang}}, \bibinfo
  {author} {\bibfnamefont {H.}~\bibnamefont {Lu}}, \bibinfo {author}
  {\bibfnamefont {Y.}~\bibnamefont {Zhang}},\ and\ \bibinfo {author}
  {\bibfnamefont {M.}~\bibnamefont {Tian}},\ }\bibfield  {title} {\bibinfo
  {title} {Thickness-tuned transition of band topology in
  $\mathrm{ZrT}{\mathrm{e}}_{5}$ nanosheets},\ }\href
  {https://doi.org/10.1103/PhysRevB.95.125135} {\bibfield  {journal} {\bibinfo
  {journal} {Phys. Rev. B}\ }\textbf {\bibinfo {volume} {95}},\ \bibinfo
  {pages} {125135} (\bibinfo {year} {2017})}\BibitemShut {NoStop}%
\bibitem [{\citenamefont {Qiu}\ \emph {et~al.}(2016)\citenamefont {Qiu},
  \citenamefont {Du}, \citenamefont {Charnas}, \citenamefont {Zhou},
  \citenamefont {Jin}, \citenamefont {Luo}, \citenamefont {Zemlyanov},
  \citenamefont {Xu}, \citenamefont {Cheng},\ and\ \citenamefont
  {Ye}}]{multiband-and-sdho_gang_ooeipahmflz_2016}%
  \BibitemOpen
  \bibfield  {author} {\bibinfo {author} {\bibfnamefont {G.}~\bibnamefont
  {Qiu}}, \bibinfo {author} {\bibfnamefont {Y.}~\bibnamefont {Du}}, \bibinfo
  {author} {\bibfnamefont {A.}~\bibnamefont {Charnas}}, \bibinfo {author}
  {\bibfnamefont {H.}~\bibnamefont {Zhou}}, \bibinfo {author} {\bibfnamefont
  {S.}~\bibnamefont {Jin}}, \bibinfo {author} {\bibfnamefont {Z.}~\bibnamefont
  {Luo}}, \bibinfo {author} {\bibfnamefont {D.~Y.}\ \bibnamefont {Zemlyanov}},
  \bibinfo {author} {\bibfnamefont {X.}~\bibnamefont {Xu}}, \bibinfo {author}
  {\bibfnamefont {G.~J.}\ \bibnamefont {Cheng}},\ and\ \bibinfo {author}
  {\bibfnamefont {P.~D.}\ \bibnamefont {Ye}},\ }\bibfield  {title} {\bibinfo
  {title} {{Observation of Optical and Electrical In-Plane Anisotropy in
  High-Mobility Few-Layer ZrTe$_5$}},\ }\href
  {https://doi.org/10.1021/acs.nanolett.6b02629} {\bibfield  {journal}
  {\bibinfo  {journal} {Nano Letters}\ }\textbf {\bibinfo {volume} {16}},\
  \bibinfo {pages} {7364} (\bibinfo {year} {2016})}\BibitemShut {NoStop}%
\bibitem [{\citenamefont {Zhang}\ \emph
  {et~al.}(2017{\natexlab{b}})\citenamefont {Zhang}, \citenamefont {Guo},
  \citenamefont {Zhu}, \citenamefont {Ma}, \citenamefont {Zheng}, \citenamefont
  {Wang}, \citenamefont {Pi}, \citenamefont {Chen}, \citenamefont {Yuan},\ and\
  \citenamefont {Tian}}]{sdho-nontrivial_zhang_dadss_2017}%
  \BibitemOpen
  \bibfield  {author} {\bibinfo {author} {\bibfnamefont {J.~L.}\ \bibnamefont
  {Zhang}}, \bibinfo {author} {\bibfnamefont {C.~Y.}\ \bibnamefont {Guo}},
  \bibinfo {author} {\bibfnamefont {X.~D.}\ \bibnamefont {Zhu}}, \bibinfo
  {author} {\bibfnamefont {L.}~\bibnamefont {Ma}}, \bibinfo {author}
  {\bibfnamefont {G.~L.}\ \bibnamefont {Zheng}}, \bibinfo {author}
  {\bibfnamefont {Y.~Q.}\ \bibnamefont {Wang}}, \bibinfo {author}
  {\bibfnamefont {L.}~\bibnamefont {Pi}}, \bibinfo {author} {\bibfnamefont
  {Y.}~\bibnamefont {Chen}}, \bibinfo {author} {\bibfnamefont {H.~Q.}\
  \bibnamefont {Yuan}},\ and\ \bibinfo {author} {\bibfnamefont {M.~L.}\
  \bibnamefont {Tian}},\ }\bibfield  {title} {\bibinfo {title} {{Disruption of
  the Accidental Dirac Semimetal State in ${\mathrm{ZrTe}}_{5}$ under
  Hydrostatic Pressure}},\ }\href
  {https://doi.org/10.1103/PhysRevLett.118.206601} {\bibfield  {journal}
  {\bibinfo  {journal} {Phys. Rev. Lett.}\ }\textbf {\bibinfo {volume} {118}},\
  \bibinfo {pages} {206601} (\bibinfo {year} {2017}{\natexlab{b}})}\BibitemShut
  {NoStop}%
\bibitem [{\citenamefont {Wang}\ \emph {et~al.}(2018)\citenamefont {Wang},
  \citenamefont {Niu}, \citenamefont {Yan}, \citenamefont {Li}, \citenamefont
  {Bi}, \citenamefont {Yao}, \citenamefont {Yu},\ and\ \citenamefont
  {Wu}}]{sdho-nontrivial_wang_vqods_2018}%
  \BibitemOpen
  \bibfield  {author} {\bibinfo {author} {\bibfnamefont {J.}~\bibnamefont
  {Wang}}, \bibinfo {author} {\bibfnamefont {J.}~\bibnamefont {Niu}}, \bibinfo
  {author} {\bibfnamefont {B.}~\bibnamefont {Yan}}, \bibinfo {author}
  {\bibfnamefont {X.}~\bibnamefont {Li}}, \bibinfo {author} {\bibfnamefont
  {R.}~\bibnamefont {Bi}}, \bibinfo {author} {\bibfnamefont {Y.}~\bibnamefont
  {Yao}}, \bibinfo {author} {\bibfnamefont {D.}~\bibnamefont {Yu}},\ and\
  \bibinfo {author} {\bibfnamefont {X.}~\bibnamefont {Wu}},\ }\bibfield
  {title} {\bibinfo {title} {{Vanishing quantum oscillations in Dirac semimetal
  ZrTe$_5$}},\ }\href {https://doi.org/10.1073/pnas.1804958115} {\bibfield
  {journal} {\bibinfo  {journal} {Proceedings of the National Academy of
  Sciences}\ }\textbf {\bibinfo {volume} {115}},\ \bibinfo {pages} {9145}
  (\bibinfo {year} {2018})}\BibitemShut {NoStop}%
\bibitem [{\citenamefont {Zhuo}\ \emph {et~al.}(2022)\citenamefont {Zhuo},
  \citenamefont {Lei}, \citenamefont {Zhu}, \citenamefont {Sun}, \citenamefont
  {Cui}, \citenamefont {Wang}, \citenamefont {Wang}, \citenamefont {Wu},
  \citenamefont {Ying}, \citenamefont {Xiang},\ and\ \citenamefont
  {Chen}}]{sdho-thick_zhuo-monolayer-tdesl2D_2022}%
  \BibitemOpen
  \bibfield  {author} {\bibinfo {author} {\bibfnamefont {W.~Z.}\ \bibnamefont
  {Zhuo}}, \bibinfo {author} {\bibfnamefont {B.}~\bibnamefont {Lei}}, \bibinfo
  {author} {\bibfnamefont {C.~S.}\ \bibnamefont {Zhu}}, \bibinfo {author}
  {\bibfnamefont {Z.~L.}\ \bibnamefont {Sun}}, \bibinfo {author} {\bibfnamefont
  {J.~H.}\ \bibnamefont {Cui}}, \bibinfo {author} {\bibfnamefont {W.~X.}\
  \bibnamefont {Wang}}, \bibinfo {author} {\bibfnamefont {Z.~Y.}\ \bibnamefont
  {Wang}}, \bibinfo {author} {\bibfnamefont {T.}~\bibnamefont {Wu}}, \bibinfo
  {author} {\bibfnamefont {J.~J.}\ \bibnamefont {Ying}}, \bibinfo {author}
  {\bibfnamefont {Z.~J.}\ \bibnamefont {Xiang}},\ and\ \bibinfo {author}
  {\bibfnamefont {X.~H.}\ \bibnamefont {Chen}},\ }\bibfield  {title} {\bibinfo
  {title} {{Thickness-dependent electronic structure in layered
  ${\mathrm{ZrTe}}_{5}$ down to the two-dimensional limit}},\ }\href
  {https://doi.org/10.1103/PhysRevB.106.085428} {\bibfield  {journal} {\bibinfo
   {journal} {Phys. Rev. B}\ }\textbf {\bibinfo {volume} {106}},\ \bibinfo
  {pages} {085428} (\bibinfo {year} {2022})}\BibitemShut {NoStop}%
\bibitem [{\citenamefont {Kamm}\ \emph {et~al.}(1985)\citenamefont {Kamm},
  \citenamefont {Gillespie}, \citenamefont {Ehrlich}, \citenamefont {Wieting},\
  and\ \citenamefont {Levy}}]{SdHO_kamm-multifreq-fsemdt_1985}%
  \BibitemOpen
  \bibfield  {author} {\bibinfo {author} {\bibfnamefont {G.~N.}\ \bibnamefont
  {Kamm}}, \bibinfo {author} {\bibfnamefont {D.~J.}\ \bibnamefont {Gillespie}},
  \bibinfo {author} {\bibfnamefont {A.~C.}\ \bibnamefont {Ehrlich}}, \bibinfo
  {author} {\bibfnamefont {T.~J.}\ \bibnamefont {Wieting}},\ and\ \bibinfo
  {author} {\bibfnamefont {F.}~\bibnamefont {Levy}},\ }\bibfield  {title}
  {\bibinfo {title} {{Fermi surface, effective masses, and Dingle temperatures
  of ${\mathrm{ZrTe}}_{5}$ as derived from the Shubnikov--de Haas effect}},\
  }\href {https://doi.org/10.1103/PhysRevB.31.7617} {\bibfield  {journal}
  {\bibinfo  {journal} {Phys. Rev. B}\ }\textbf {\bibinfo {volume} {31}},\
  \bibinfo {pages} {7617} (\bibinfo {year} {1985})}\BibitemShut {NoStop}%
\bibitem [{\citenamefont {Chi}\ \emph {et~al.}(2017)\citenamefont {Chi},
  \citenamefont {Zhang}, \citenamefont {Gu}, \citenamefont {Kharzeev},
  \citenamefont {Dai},\ and\ \citenamefont
  {Li}}]{multicar-secondband_chi-lifshitz-ltmetab_2017}%
  \BibitemOpen
  \bibfield  {author} {\bibinfo {author} {\bibfnamefont {H.}~\bibnamefont
  {Chi}}, \bibinfo {author} {\bibfnamefont {C.}~\bibnamefont {Zhang}}, \bibinfo
  {author} {\bibfnamefont {G.}~\bibnamefont {Gu}}, \bibinfo {author}
  {\bibfnamefont {D.~E.}\ \bibnamefont {Kharzeev}}, \bibinfo {author}
  {\bibfnamefont {X.}~\bibnamefont {Dai}},\ and\ \bibinfo {author}
  {\bibfnamefont {Q.}~\bibnamefont {Li}},\ }\bibfield  {title} {\bibinfo
  {title} {{Lifshitz transition mediated electronic transport anomaly in bulk
  ZrTe$_5$}},\ }\href {https://doi.org/10.1088/1367-2630/aa55a3} {\bibfield
  {journal} {\bibinfo  {journal} {New Journal of Physics}\ }\textbf {\bibinfo
  {volume} {19}},\ \bibinfo {pages} {015005} (\bibinfo {year}
  {2017})}\BibitemShut {NoStop}%
\bibitem [{\citenamefont {Niu}\ \emph {et~al.}(2017)\citenamefont {Niu},
  \citenamefont {Wang}, \citenamefont {He}, \citenamefont {Zhang},
  \citenamefont {Li}, \citenamefont {Cai}, \citenamefont {Ma}, \citenamefont
  {Jia}, \citenamefont {Yu},\ and\ \citenamefont
  {Wu}}]{multiband-nano_jj-etnzs_2017}%
  \BibitemOpen
  \bibfield  {author} {\bibinfo {author} {\bibfnamefont {J.}~\bibnamefont
  {Niu}}, \bibinfo {author} {\bibfnamefont {J.}~\bibnamefont {Wang}}, \bibinfo
  {author} {\bibfnamefont {Z.}~\bibnamefont {He}}, \bibinfo {author}
  {\bibfnamefont {C.}~\bibnamefont {Zhang}}, \bibinfo {author} {\bibfnamefont
  {X.}~\bibnamefont {Li}}, \bibinfo {author} {\bibfnamefont {T.}~\bibnamefont
  {Cai}}, \bibinfo {author} {\bibfnamefont {X.}~\bibnamefont {Ma}}, \bibinfo
  {author} {\bibfnamefont {S.}~\bibnamefont {Jia}}, \bibinfo {author}
  {\bibfnamefont {D.}~\bibnamefont {Yu}},\ and\ \bibinfo {author}
  {\bibfnamefont {X.}~\bibnamefont {Wu}},\ }\bibfield  {title} {\bibinfo
  {title} {{Electrical transport in nanothick ${\mathrm{ZrTe}}_{5}$ sheets:
  From three to two dimensions}},\ }\href
  {https://doi.org/10.1103/PhysRevB.95.035420} {\bibfield  {journal} {\bibinfo
  {journal} {Phys. Rev. B}\ }\textbf {\bibinfo {volume} {95}},\ \bibinfo
  {pages} {035420} (\bibinfo {year} {2017})}\BibitemShut {NoStop}%
\bibitem [{\citenamefont {Yang}\ \emph {et~al.}(2019)\citenamefont {Yang},
  \citenamefont {Wang}, \citenamefont {Zhang}, \citenamefont {Wang},
  \citenamefont {He}, \citenamefont {Liu},\ and\ \citenamefont
  {Xu}}]{quasi2D_yang-quantOsc-qonsq2dds-2019}%
  \BibitemOpen
  \bibfield  {author} {\bibinfo {author} {\bibfnamefont {P.}~\bibnamefont
  {Yang}}, \bibinfo {author} {\bibfnamefont {W.}~\bibnamefont {Wang}}, \bibinfo
  {author} {\bibfnamefont {X.}~\bibnamefont {Zhang}}, \bibinfo {author}
  {\bibfnamefont {K.}~\bibnamefont {Wang}}, \bibinfo {author} {\bibfnamefont
  {L.}~\bibnamefont {He}}, \bibinfo {author} {\bibfnamefont {W.}~\bibnamefont
  {Liu}},\ and\ \bibinfo {author} {\bibfnamefont {Y.}~\bibnamefont {Xu}},\
  }\bibfield  {title} {\bibinfo {title} {{Quantum Oscillations from Nontrivial
  States in Quasi-Two-Dimensional Dirac Semimetal ZrTe$_5$ Nanowires}},\ }\href
  {https://doi.org/10.1038/s41598-019-39144-y} {\bibfield  {journal} {\bibinfo
  {journal} {Scientific Reports}\ }\textbf {\bibinfo {volume} {9}},\ \bibinfo
  {pages} {3558} (\bibinfo {year} {2019})}\BibitemShut {NoStop}%
\bibitem [{\citenamefont {Tajkov}\ \emph {et~al.}(2022)\citenamefont {Tajkov},
  \citenamefont {Nagy}, \citenamefont {Kandrai}, \citenamefont {Koltai},
  \citenamefont {Oroszl{\'a}ny}, \citenamefont {S{\"u}le}, \citenamefont
  {Horv{\'a}th}, \citenamefont {Vancs{\'o}}, \citenamefont {Tapaszt{\'o}},\
  and\ \citenamefont {Nemes-Incze}}]{topo_tajkov-phase-strain_2022}%
  \BibitemOpen
  \bibfield  {author} {\bibinfo {author} {\bibfnamefont {Z.}~\bibnamefont
  {Tajkov}}, \bibinfo {author} {\bibfnamefont {D.}~\bibnamefont {Nagy}},
  \bibinfo {author} {\bibfnamefont {K.}~\bibnamefont {Kandrai}}, \bibinfo
  {author} {\bibfnamefont {J.}~\bibnamefont {Koltai}}, \bibinfo {author}
  {\bibfnamefont {L.}~\bibnamefont {Oroszl{\'a}ny}}, \bibinfo {author}
  {\bibfnamefont {P.}~\bibnamefont {S{\"u}le}}, \bibinfo {author}
  {\bibfnamefont {Z.~E.}\ \bibnamefont {Horv{\'a}th}}, \bibinfo {author}
  {\bibfnamefont {P.}~\bibnamefont {Vancs{\'o}}}, \bibinfo {author}
  {\bibfnamefont {L.}~\bibnamefont {Tapaszt{\'o}}},\ and\ \bibinfo {author}
  {\bibfnamefont {P.}~\bibnamefont {Nemes-Incze}},\ }\bibfield  {title}
  {\bibinfo {title} {{Revealing the topological phase diagram of ZrTe$_5$ using
  the complex strain fields of microbubbles}},\ }\href
  {https://doi.org/10.1038/s41524-022-00854-z} {\bibfield  {journal} {\bibinfo
  {journal} {npj Computational Materials}\ }\textbf {\bibinfo {volume} {8}},\
  \bibinfo {pages} {177} (\bibinfo {year} {2022})}\BibitemShut {NoStop}%
\bibitem [{\citenamefont {Shen}\ \emph {et~al.}(2017)\citenamefont {Shen},
  \citenamefont {Wang}, \citenamefont {Sun}, \citenamefont {Jiang},
  \citenamefont {Xu}, \citenamefont {Zhang}, \citenamefont {Zhang},
  \citenamefont {Lv}, \citenamefont {Yao}, \citenamefont {Chen}, \citenamefont
  {Lu}, \citenamefont {Chen}, \citenamefont {Felser}, \citenamefont {Yan},
  \citenamefont {Liu}, \citenamefont {Yang},\ and\ \citenamefont
  {Chen}}]{sideband-epocket_shen-segesb_2017}%
  \BibitemOpen
  \bibfield  {author} {\bibinfo {author} {\bibfnamefont {L.}~\bibnamefont
  {Shen}}, \bibinfo {author} {\bibfnamefont {M.}~\bibnamefont {Wang}}, \bibinfo
  {author} {\bibfnamefont {S.}~\bibnamefont {Sun}}, \bibinfo {author}
  {\bibfnamefont {J.}~\bibnamefont {Jiang}}, \bibinfo {author} {\bibfnamefont
  {X.}~\bibnamefont {Xu}}, \bibinfo {author} {\bibfnamefont {T.}~\bibnamefont
  {Zhang}}, \bibinfo {author} {\bibfnamefont {Q.}~\bibnamefont {Zhang}},
  \bibinfo {author} {\bibfnamefont {Y.}~\bibnamefont {Lv}}, \bibinfo {author}
  {\bibfnamefont {S.}~\bibnamefont {Yao}}, \bibinfo {author} {\bibfnamefont
  {Y.}~\bibnamefont {Chen}}, \bibinfo {author} {\bibfnamefont {M.}~\bibnamefont
  {Lu}}, \bibinfo {author} {\bibfnamefont {Y.}~\bibnamefont {Chen}}, \bibinfo
  {author} {\bibfnamefont {C.}~\bibnamefont {Felser}}, \bibinfo {author}
  {\bibfnamefont {B.}~\bibnamefont {Yan}}, \bibinfo {author} {\bibfnamefont
  {Z.}~\bibnamefont {Liu}}, \bibinfo {author} {\bibfnamefont {L.}~\bibnamefont
  {Yang}},\ and\ \bibinfo {author} {\bibfnamefont {Y.}~\bibnamefont {Chen}},\
  }\bibfield  {title} {\bibinfo {title} {{Spectroscopic evidence for the
  gapless electronic structure in bulk ZrTe$_5$}},\ }\href
  {https://doi.org/10.1016/j.elspec.2016.10.007} {\bibfield  {journal}
  {\bibinfo  {journal} {Journal of Electron Spectroscopy and Related
  Phenomena}\ }\textbf {\bibinfo {volume} {219}},\ \bibinfo {pages} {45}
  (\bibinfo {year} {2017})}\BibitemShut {NoStop}%
\bibitem [{\citenamefont {Fülöp}\ \emph {et~al.}(2021)\citenamefont
  {Fülöp}, \citenamefont {M{\'{a}}rffy}, \citenamefont {T{\'{o}}v{\'{a}}ri},
  \citenamefont {Kedves}, \citenamefont {Zihlmann}, \citenamefont {Indolese},
  \citenamefont {Kov{\'{a}}cs-Krausz}, \citenamefont {Watanabe}, \citenamefont
  {Taniguchi}, \citenamefont {Schönenberger}, \citenamefont
  {K{\'{e}}zsm{\'{a}}rki}, \citenamefont {Makk},\ and\ \citenamefont
  {Csonka}}]{pressure_fulop-vdW-pressure_2021}%
  \BibitemOpen
  \bibfield  {author} {\bibinfo {author} {\bibfnamefont {B.}~\bibnamefont
  {Fülöp}}, \bibinfo {author} {\bibfnamefont {A.}~\bibnamefont
  {M{\'{a}}rffy}}, \bibinfo {author} {\bibfnamefont {E.}~\bibnamefont
  {T{\'{o}}v{\'{a}}ri}}, \bibinfo {author} {\bibfnamefont {M.}~\bibnamefont
  {Kedves}}, \bibinfo {author} {\bibfnamefont {S.}~\bibnamefont {Zihlmann}},
  \bibinfo {author} {\bibfnamefont {D.}~\bibnamefont {Indolese}}, \bibinfo
  {author} {\bibfnamefont {Z.}~\bibnamefont {Kov{\'{a}}cs-Krausz}}, \bibinfo
  {author} {\bibfnamefont {K.}~\bibnamefont {Watanabe}}, \bibinfo {author}
  {\bibfnamefont {T.}~\bibnamefont {Taniguchi}}, \bibinfo {author}
  {\bibfnamefont {C.}~\bibnamefont {Schönenberger}}, \bibinfo {author}
  {\bibfnamefont {I.}~\bibnamefont {K{\'{e}}zsm{\'{a}}rki}}, \bibinfo {author}
  {\bibfnamefont {P.}~\bibnamefont {Makk}},\ and\ \bibinfo {author}
  {\bibfnamefont {S.}~\bibnamefont {Csonka}},\ }\bibfield  {title} {\bibinfo
  {title} {{New method of transport measurements on van der Waals
  heterostructures under pressure}},\ }\href
  {https://doi.org/10.1063/5.0058583} {\bibfield  {journal} {\bibinfo
  {journal} {Journal of Applied Physics}\ }\textbf {\bibinfo {volume} {130}},\
  \bibinfo {pages} {064303} (\bibinfo {year} {2021})}\BibitemShut {NoStop}%
\end{thebibliography}
%

\end{document}